\documentclass[aps,prb,twocolumn,superscriptaddress,showpacs,amssymb,amsmath]{revtex4}

\usepackage{graphicx}
\newcommand{\z}{\text {z}}
\usepackage{units}
\usepackage{txfonts}
\begin{document}
\title{Charge and spin diffusion on the metallic side of the metal-insulator transition: a self-consistent approach}

\author{Thomas Wellens}
\affiliation{Physikalisches Institut der Albert-Ludwigs-Universität, Hermann-Herder-Str. 3, D-79104 Freiburg, Germany}

\author{Rodolfo A.\ Jalabert}
\affiliation{Freiburg Institute for Advanced Studies, Albert-Ludwigs-Universit\"at, Albertstr.\ 19, D-79104 Freiburg, Germany}
\affiliation{Institut de Physique et Chimie des Mat\'{e}riaux de Strasbourg, Universit\'e de Strasbourg, CNRS UMR 7504, 23 rue du Loess, BP 43, 67034 Strasbourg Cedex 2, France}

\begin{abstract}
We develop a self-consistent theory describing the spin and spatial electron diffusion in the impurity band of doped semiconductors under the effect of a weak spin-orbit coupling. The resulting low-temperature spin-relaxation time and diffusion coefficient are calculated within different schemes of the self-consistent framework. The simplest of these schemes qualitatively reproduces previous phenomenological developments,
while more elaborate calculations provide corrections that approach the values obtained in numerical simulations. The results are universal for zinc-blende semiconductors with electron conductance in the impurity band, and thus they are able to account for the measured spin-relaxation times of materials with very different physical parameters. From a general point of view, our theory opens a new perspective for describing the hopping dynamics in random quantum networks.
\end{abstract}

\maketitle

\section{Introduction}
\label{sec:intro}

The low-temperature spin-relaxation time ($\tau_\mathrm{s}$) of n-doped
semiconductors presents a maximum for doping densities near that of the metal-insulator transition (MIT) \cite{cha,zar-cas,kik-aws,dzh,oes-roe-hau-hae,sch-hei-roh,roe-ber-mue,spr-etal}. This experimental observation is particularly intriguing as it encompasses both, the rich physics of spin-orbit coupling in semiconductors and a paradigmatic quantum phase transition. On the one hand, the spin dynamics in different semiconductor-based systems is relevant from the fundamental point of view, as well as for potential applications of spintronics and quantum information technologies \cite{fab-etal,wu-jia-wen}. On the other hand, despite a substantial research effort, the MIT remains one of the most challenging open problems in condensed matter physics \cite{vLoe,gehard,slevin14}.

While the mechanisms behind spin relaxation were promptly identified for the regime of high temperatures or for doping densities far away from the critical one \cite{dzh,lau01,jia-wu}, the understanding of the low-temperature spin-relaxation close to the MIT required a sustained theoretical effort \cite{shk,kav,put-joy,tam-wei-jal,prl2012}. On the metallic side of the transition (for impurity densities $n_\mathrm{i}$ slightly larger than the critical one $n_\mathrm{c}$) the Dresselhaus spin-orbit coupling was identified as the source of the spin-relaxation in the case of zinc-blende semiconductors. In particular, it was proposed that when the electron conduction is in the impurity band ($n_\mathrm{c}\!<\!n_\mathrm{i}\!<\!n_\mathrm{h}$), the spin-relaxation rate is given by \cite{prl2012,twj2016}
\begin{equation}\label{eq:taus_vs_dens}
\frac{1}{\tau_\mathrm{s}} = 0.36 \ \frac{\gamma^2}{a^6V_0\hbar} \, 
\mathcal{N}_\mathrm{i}^{1/2}. 
\end{equation}
$\gamma$ is the material-dependent Dresselhaus coupling constant 
\cite{car-chr-fas,fab-etal,eng-ras-hal,win}, while the other parameters in 
Eq. \eqref{eq:taus_vs_dens} depend on the nature of the impurity states: the effective Bohr radius ($a$), the binding energy ($V_0/2)$, and the dimensionless impurity density ($\mathcal{N}_\mathrm{i}=n_\mathrm{i}a^3$). The hybridization density $n_\mathrm{h}$ marks the impurity concentration beyond which there is a considerable overlap between the impurity and conduction bands. 

The form \eqref{eq:taus_vs_dens} of the relaxation rate is quite general. In the above-specified density interval, it applies to any zinc-blende semiconductor, with the possible exception of narrow-gap materials. Indeed, it has been shown to give good account, within the experimental uncertainties and the limited knowledge of some material parameters, of the spin-relaxation measured in GaAs and CdTe, despite the very different material constants of these two semiconductors. For instance, the Mott criterion for the MIT setting the critical dimensionless impurity density $\mathcal{N}_\mathrm{c}=n_\mathrm{c}a^3\simeq 0.017$,
leads to $n_\mathrm{c}=\unit[2\times 10^{16}]{cm^{-3}}$ ($\unit[9\times 10^{16}]{cm^{-3}}$) for GaAs (CdTe), and there are two orders of magnitude difference between the corresponding values of $\tau_\mathrm{s}$ for these two cases\cite{kik-aws,dzh,oes-roe-hau-hae,sch-hei-roh,roe-ber-mue,spr-etal}.

Eq. \eqref{eq:taus_vs_dens} has been analytically derived and it is in good agreement with numerical simulations \cite{prl2012}. Both theoretical approaches (analytic and numerical) are based on a generalization of the well-known Matsubara-Toyozawa (MT) model \cite{mat-toy}, describing the diffusion of non-interacting electrons  through randomly distributed impurity sites, so as to include spin-flipping hopping terms \cite{tam-wei-jal,proceedings}. In the numerical approach, the spin-relaxation time is extracted from the evolution of  
initial states with
a well-defined spin projection. 
The weakness of the spin-orbit coupling and the finite system sizes that can be handled require the use of delicate extrapolations and a finite-size scaling analysis. 
In the analytic formulation, the spin-relaxation rate is obtained from the diffusive accumulation of spin rotation angles as the electron jumps between impurity centers. Such a phenomenological approach needs to be put on a firm basis as a well-controlled approximation that can be extended in a systematic way in order to accurately describe various parameter regimes.

The present work addresses the above-mentioned task and develops a systematic self-consistent diagrammatic perturbation approach to obtain the long-time charge and spin dynamics in a disordered network of impurity sites. The phenomenological result of Eq. \eqref{eq:taus_vs_dens} is qualitatively recovered using a simple 
self-consistent approximation in a locator expansion of the self-energy which can be analytically solved.
In a second approximation, we include diagrams describing loops   of arbitrary length, which can be shown to give the dominant contributions in the limit of high impurity densities \cite{mat-toy}. The resulting charge diffusion coincides with that of the MT diagrammatic expansion\cite{mat-toy}, and the spin-relaxation rate improves the previous estimation,  correcting its density dependence. 

Further refinements of the self-consistent approach can be implemented in a systematic way by considering repeated scattering 
with given impurities (i.e. cross diagrams). In particular, we consider repeated scattering from pairs of impurities, which has been shown to be dominant for low-concentration densities \cite{Elyutin}. We show that these processes lead to small corrections for the range of impurity densities we are interested in ($n_\mathrm{c}\!<\!n_\mathrm{i}\!<\!n_\mathrm{h}$),
thus confirming the applicability of the high-density limit in this regime.

Our approach relies on two drastic simplifications: working with a zero-temperature formalism and ignoring electron-electron interaction. Detailed temperature-dependent measurements of the spin-relaxation time \cite{roe-ber-mue} yielded a saturation of $\tau_\mathrm{s}$ below \unit[10]{K}, indicating that inelastic processes can be ignored at low temperatures. In the case of the n-doped semiconductors, the MIT appears at a doping density $n_\mathrm{c}$ where the Fermi level is in the impurity band \cite{Shklovskii,Economou,Romero}. Electron-electron interactions are crucial in order to account for the observed features of the MIT \cite{vLoe,gehard,slevin14} and induce significant many-body effects on the insulating side of the transition ($n_\mathrm{i} < n_\mathrm{c}$). Therefore, we restrict our work to metallic densities ($n_\mathrm{i} > n_\mathrm{c}$) away from the critical region, where a single-particle description is possible.  

Our main interest is in the extension of the self-consistent approach in order to include spin-orbit effects in the MT model of the impurity band, leading to the zero-temperature density-dependent spin-relaxation time. Additionally, the treatment of the high and low impurity concentrations within the same self-consistent framework constitutes a useful development for the much studied spinless MT model \cite{mat-toy,Elyutin,yonezawa64,mat-kan,gas-cyr,chao-oli-majlis,pur-oda_81,gibbons_81,chi-hub} and other cases where a resonant excitation is able to jump between the sites of a disordered network. In general, our approach can be applied to any random network model with hopping matrix elements depending on the distances between randomly placed sites.
The problem of the excitation transport in ultra cold Rydberg gases is an example of such a disordered network, which has recently received considerable attention experimentally \cite{Rydberg_ex} and theoretically \cite{Rydberg_th}. Another example is provided by molecular light-harvesting complexes \cite{amerongen00,scholak11} (at least for times shorter than the decoherence times induced by coupling  to environmental degrees of freedom), where the impact of quantum coherence on excitation transport is under current debate \cite{engel07}.

The paper is organized as follows. In Sec.~\ref{sec:socib} we generalize the MT model by including the spin-orbit coupling through a matrix of hopping amplitudes. The transformation symmetries of this matrix, thoroughly exploited throughout our work, are established. In Sec.~\ref{sec:pqd} we lay the basis for the calculation of the quantum evolution of the orbital and spin degrees of freedom, introducing the main definitions used in this article. Sec.~\ref{sec:le} presents the locator expansion in its matrix form and establishes the general properties fulfilled by the Green function and the self-energy. The self-consistent theory is developed in Sec.~\ref{sec:sca}, using the Bethe-Salpeter equation and the Ward identity in order to extract the long-time orbital and spin dynamics. Secs.~\ref{sec:simple}, \ref{sec:lcsca}, and \ref{sec:scasd} tackle three levels of approximation of increasing complexity within the self-consistent scheme. Results for the density of states, the spatial diffusion coefficient and the spin-relaxation rate are obtained and discussed. We provide conclusions in Sec.~\ref{sec:conclusion}, and we relegate to the appendices the derivation of the charge and spin dynamics from the intensity propagator, the technical aspects of the self-consistent approximation in our random impurity model with spin-orbit interaction, as well as the calculation of some auxiliary quantities. 

\section{Spin-orbit coupling in the impurity band}
\label{sec:socib}

\subsection{Hamiltonian and hopping amplitude matrix}
\label{subsec:ha}

The envelope-function approximation for electrons in the conduction-band of zinc blende semiconductors incorporates the 
crystal lattice-scale physics into
the effective one-body Hamiltonian \cite{noz-lew,eng-ras-hal}
\begin{subequations}
\label{eq:Hallunrestr}
\begin{eqnarray} 
H &=& H_0 + H_{\text{D}} + H_{\text{extr}} \, ,
\label{eq:Htot}
\\ 
H_0 &=& \frac{p^2}{2 \, m^*} + V(\mathbf{r}) \, ,
\label{eq:Hzero}
\\ 
H_{\text{D}} &=& \gamma \, [\sigma_x k_x (k_y^2 - k_z^2) + 
                          \text{cyclic permutations}] \, , 
\label{eq:BIA}
\\
H_{\text{extr}} &=& \lambda \, \boldsymbol{\sigma} \cdot \nabla V 
                 \times \mathbf{k} \, .
\label{eq:SIA}
\end{eqnarray}
\end{subequations}
The spin-independent part $H_0$ is determined by the effective mass ($m^*$) and the electrostatic potential $V(\mathbf{r})$ including all potentials aside from that of the crystal lattice. We note $\mathbf{p}$ the momentum operator, $\mathbf{k} = \mathbf{p}/\hbar$, and $\boldsymbol{\sigma}$ the vector of Pauli matrices. The Dresselhaus (intrinsic) term $H_{\text{D}}$ is enabled by the bulk inversion asymmetry. Typically \cite{win}, $\gamma = 27$ ${{\rm eV}\AA^3}$ ($44$ ${{\rm eV}\AA^3}$) is used for GaAs (CdTe). However, the precise value of this coupling constant is a matter of current debate 
\cite{eng-ras-hal,fab-etal,car-chr-fas,cha-van-kot,kri-hal,mar-ste-tit,knap-etal,mei-etal,jusserand}. The extrinsic term $H_{\text{extr}}$ has the same form as the spin-orbit interaction in vacuum, but the effective coupling constant $\lambda$ is usually orders of magnitude larger than the vacuum one ($\lambda_0 = \unit[3.7 \times 10^{-6}]{\AA^2}$). 
Nevertheless, as argued below, this term turns out to be irrelevant in comparison to the extrinsic term $H_{\text{D}}$ for the problem of spin relaxation.

For n-doped semiconductors 
\begin{equation} 
V(\mathbf{r}) = \sum_m V_m(\mathbf{r}) \, ,
\end{equation}
where
\begin{equation}
V_m(\mathbf{r}) = - \ \frac{e^2}{\epsilon |\mathbf{r}-\mathbf{R}_m|}
\label{eq:vindpot}
\end{equation}
is the hydrogenic potential of an impurity placed at $\mathbf{R}_m$, and $\epsilon$ is the dielectric constant of the semiconductor. A possible refinement of our single-particle approach, not pursued in this work, is to trade the potential \eqref{eq:vindpot} by a Thomas-Fermi effective potential that includes the effect of screening \cite{Serre}. 

The energy of the electronic ground state of an isolated impurity ($\varepsilon_{00}$) is $V_0/2$ below the bottom of the conduction band ($V_0=e^2 /\epsilon a$). The corresponding electronic wavefunction for impurity $m$ reads $\phi_m(\mathbf{r})= \phi(|\mathbf{r}-\mathbf{R}_m|)$, with $\phi(\mathbf{r}) = (1/\sqrt{\pi a^3}) \exp{(-r/a)}$.  

The overlap between the impurity centers $m=1,\ldots,N$ widens their energy levels into an impurity band \cite{mat-toy,yonezawa64,mat-kan,gas-cyr,chao-oli-majlis,pur-oda_81,chi-hub,Serre,hal-lax}. Combined magnetotransport and far-infrared spectroscopy allow to probe the impurity band and demonstrate that in the vicinity of the MIT the electrons are confined to such a band \cite{Romero}. Henceforth, we therefore restrict our efforts to describing the charge and spin dynamics of electrons in the impurity band.

The electronic ground states of the isolated impurity sites $m$ provide a restricted basis $\{| m \sigma \rangle\}$ to describe the electron jumping between impurity centers ($\sigma=\pm$ corresponds to a spin projection of $\pm \hbar/2$ in the $z$-direction). 
The Hamiltonian in this restricted space can be expressed as 
\begin{subequations}
\label{eq:Hall}
\begin{eqnarray} 
{\cal H} &=& {\cal H}_{00} + {\cal H}_{\text{c}} \, ,
\label{eq:Hrest}
\\ 
{\cal H}_{00} &=& \sum_{m} \sum_{\sigma}   
	| m \sigma\rangle \ \varepsilon_{00} \ \langle m \sigma | \, ,
\label{eq:Hzerozero}
\\ 
{\cal H}_{\text{c}} &=& \sum_{m' \neq m} \sum_{\sigma' \sigma} 
	| m' \sigma'\rangle  {\mathcal V}^{\sigma',\sigma}({\bf R}_{m'm}) \langle m \sigma | \ .
\label{eq:coupling}
\end{eqnarray}
\end{subequations}
Choosing $\varepsilon_{00}$ as the energy origin, we can ignore the first term ${\cal H}_{00}$. The coupling matrix elements (hopping amplitudes) 
can be obtained from the original (i.e. unrestricted) Hamiltonian $H$ of Eq.~\eqref{eq:Hallunrestr} as  
\begin{equation}
 {\mathcal V}^{\sigma',\sigma}({\bf R}_{m'm})=
 \langle m'\sigma'|H|m\sigma\rangle \, .
\label{eq:H}
\end{equation}
Due to the crystal translational symmetry, these matrix elements only depend on the relative position ${\bf R}_{m'm}={\bf R}_{m'}-{\bf R}_{m}$ between the two substituting impurities at sites $m'\neq m$. 

The spin-independent hopping amplitudes are
\begin{eqnarray} 
{\mathcal V}_{0}({\bf R}_{m'm}) &=& \langle m' \sigma | H_0 | m \sigma\rangle
\simeq \langle m'\sigma | V_{m'} | m \sigma \rangle
\, .
\label{eq:meh00}
\end{eqnarray}
We have done the standard approximation of neglecting the three-center integrals $\langle m' \sigma | V_p | m \sigma\rangle$ with $p \neq m,m'$. The exponential decay of $\phi(\mathbf{r})$ makes these terms much smaller than the two-center integral $\langle m'\sigma | V_{m'} | m \sigma \rangle$, resulting in
\begin{equation} 
{\mathcal V}_{0}({\bf r}) =  -V_0 \left(1+\frac{r}{a} \right) \ e^{-r/a} 
\, .
\label{eq:meh0}
\end{equation}
We note ${\bf r}=(x,y,z)$ and $r=|{\bf r}|$. 
The matrix elements \eqref{eq:meh0} define the Matsubara-Toyozawa model \cite{mat-toy}.
The subtleties, drawbacks and applicability of this model to describe the 
metallic side of the MIT, have been extensively studied \cite{yonezawa64,mat-kan,chi-hub,proceedings}. 

The spin-dependent hopping amplitudes have two contributions coming, respectively, from $H_{\text{D}}$ and $H_{\text{extr}}$. For the latter, it has been shown that the non-vanishing terms arise from three-center integrals \cite{tam-wei-jal}, resulting in an extremely small contribution to the spin-mixing matrix element. Therefore, we only keep the Dresselhaus contribution 
\begin{equation}
{\mathcal V}_{\text{D}}^{\sigma',\sigma}({\bf R}_{m'm}) = 
\langle m' \sigma' | H_{\text{D}} | m \sigma \rangle \, .
\end{equation}
Noting $\bar{\sigma}=-\sigma$, the spin-flipping hopping amplitudes and the spin-dependent contribution to the spin-conserving hopping amplitudes, respectively, write \cite{prl2012}
\begin{subequations} 
\begin{eqnarray}
\label{eq:matelbia}
{\mathcal V}_{\text{D}}^{\bar{\sigma},\sigma}({\bf r}) &=& 
	i \ {\mathcal C}_{x}({\bf r}) - \sigma \ {\mathcal C}_{y}({\bf r})
 \, , \\
{\mathcal V}_{\text{D}}^{\sigma,\sigma}({\bf r}) &=& 
    i \ \sigma \ {\mathcal C}_{z}({\bf r})  \, ,
\end{eqnarray}
\end{subequations}
where
\begin{subequations}
\begin{eqnarray}
{\mathcal C}_x({\bf r}) & = & - \frac{\gamma}{3a^5 r} \ x \ \left(y^2-z^2\right) \ e^{-r/a} \, , \\
{\mathcal C}_y({\bf r}) & = & - \frac{\gamma}{3a^5 r} \ y \ \left(z^2-x^2\right) \ e^{-r/a} \, , \\
{\mathcal C}_z({\bf r}) & = & - \frac{\gamma}{3a^5 r} \ z \ \left(x^2-y^2\right) \ e^{-r/a} \, .
\end{eqnarray}
\label{eq:Cxyz}
\end{subequations}

The hopping amplitudes \eqref{eq:H} can then be expressed through a $2\times 2$ matrix in the spin subspace as

\begin{equation}
{\mathcal V}({\bf r})=
\left(\begin{array}{cc} {\mathcal V}_0({\bf r})+i{\mathcal C}_{z}({\bf r}) & \,
i \ {\mathcal C}_{x}({\bf r}) + {\mathcal C}_{y}({\bf r}) \\
i \ {\mathcal C}_{x}({\bf r}) - {\mathcal C}_{y}({\bf r}) & \, {\mathcal V}_0({\bf r})-i{\mathcal C}_{z}({\bf r}) \end{array}\right)
\label{eq:vmatrix} \, .
\end{equation}

Since $\left|{\mathcal C}_{z}({\bf r})\right| \ll \left|{\mathcal V}_0({\bf r})\right|$ the spin-dependent hopping amplitude 
${\mathcal V}_{\text{D}}^{\sigma,\sigma}({\bf r})$ is generally omitted for the calculation of the spin-relaxation rate \cite{prl2012}. Nevertheless, this contribution is needed in order to keep the symmetries of the problem and to obtain the correct expression of the spin decoherence rate.  

The Fourier transform of ${\mathcal V}({\bf r})$ is defined by
\begin{eqnarray}
\tilde{\mathcal V}({\bf k}) &=& 
\int{\rm d}{\bf r}~e^{i{\bf k}\cdot{\bf r}}{\mathcal V}({\bf r}) 
\nonumber \\
&=&
\left(\begin{array}{cc} \tilde{\mathcal V}_0({\bf k}) + i \tilde{\mathcal C}_z({\bf k})& i \tilde{\mathcal C}_x({\bf k})+\tilde{\mathcal C}_y({\bf k}) \\
i \tilde{\mathcal C}_x({\bf k})-\tilde{\mathcal C}_y({\bf k}) & \tilde{\mathcal V}_0({\bf k}) -i \tilde{\mathcal C}_z({\bf k})\end{array}\right) \, .
\label{eq:vtilde}
\end{eqnarray}
According to \eqref{eq:meh0} and \eqref{eq:Cxyz},
\begin{subequations}
\label{eq:fth}
\begin{eqnarray}
\label{eq:vk}
\tilde{\mathcal V}_0({\bf k}) & = & - \ \frac{32 a^3 \pi V_0}{\left[1+(ka)^2\right]^3} \, , \\
\tilde{\mathcal C}_x({\bf k}) & = &   \ \frac{64 \pi i \gamma a^3}{\left[1+(ka)^2\right]^4} \ k_x(k_y^2-k_z^2) \, ,
\label{eq:cxk}
\end{eqnarray}
\end{subequations}

\noindent where ${\bf k}=(k_x,k_y,k_z)$ and  $k=|{\bf k}|$. The corresponding expressions for $\tilde{\mathcal C}_y({\bf k})$ and $\tilde{\mathcal C}_z({\bf k})$ are obtained from that of Eq.~\eqref{eq:cxk} for $\tilde{\mathcal C}_x({\bf k})$ by cyclic permutation of the spatial indices.

\subsection{Symmetries of the hopping amplitude matrix}
\label{subsec:sha}

The underlying symmetries of the zinc-blende crystal structure induce the transformation properties of the matrix ${\mathcal V}({\bf r})$. For instance, the $c_{2}$ rotations around the Cartesian axes imply that 

\begin{eqnarray}
\label{eq:symmetry2}
{\mathcal V}(x,y,z) &=& 
D_{x}^{(1/2)}(\pi) \ {\mathcal V}(x,-y,-z) \ D_{x}^{(1/2)\dagger}(\pi) 
 \nonumber \\ 
&=& 
D_{y}^{(1/2)}(\pi) \ {\mathcal V}(-x,y,-z) \ D_{y}^{(1/2)\dagger}(\pi) 
 \nonumber \\ 
&=& 
D_{z}^{(1/2)}(\pi) \ {\mathcal V}(-x,-y,z) \ D_{z}^{(1/2)\dagger}(\pi) 
 \, ,
\end{eqnarray}
where $D_{\mu}^{(1/2)}(\pi)= -i \sigma_{\mu}$ is the spin-rotation matrix of an angle of $\pi$ along the axis $\mu=(x,y,z)$.
In addition, the symmetry with respect to cyclic permutations of the axis labels ($\{x,y,z\} \to \{y,z,x\}$, and $\{\sigma_x,\sigma_y,\sigma_z\} \to \{\sigma_y,\sigma_z,\sigma_x\}$) leads to
\begin{equation}
{\mathcal V}(x,y,z)  =  P \ {\mathcal V}(y,z,x) \ P^\dagger
\, ,
\label{eq:symmv4}
\end{equation}
where
\begin{equation}
P=\frac{1}{\sqrt{2}}\left(\begin{array}{cc} 1 & -i \\ 1 & i\end{array}\right) \, .
\end{equation}

It will be later useful to establish the transformation properties of the matrices ${\mathcal V}({\bf r})$ and $\tilde{\mathcal V}({\bf k})$ under spatial inversion. The transformation ${\bf r} \to -{\bf r}$ is not a symmetry of the zinc-blende structure. While the spin, being an angular moment, is invariant under spatial inversion, the orbital part is changed according to ${\mathcal V}_{0}(-{\bf r})={\mathcal V}_{0}({\bf r})$ and ${\mathcal C}_{\mu}(-{\bf r})= - {\mathcal C}_{\mu}({\bf r})$. Similarly, $\tilde{\mathcal V}_{0}(-{\bf k})=\tilde{\mathcal V}_{0}({\bf k})$ and $\tilde{\mathcal C}_{\mu}(-{\bf k})= - \tilde{\mathcal C}_{\mu}({\bf k})$.
Therefore, 
both ${\mathcal V}({\bf r})$ and $\tilde{\mathcal V}({\bf k})$, fulfill the following property (that we refer to as 'para-odd'): under the operation of spatial inversion, the two diagonal matrix elements are interchanged and the two off-diagonal matrix elements change their sign. It is easy to show that if a $2\times 2$ matrix is para-odd, any integer power of it inherits this property. Moreover, the product of a para-odd matrix times the one obtained upon space inversion results in a diagonal matrix proportional to the $2\times 2$ identity matrix $\mathbb{I}_{2}$. For instance,
\begin{subequations}
\begin{eqnarray}
{\mathcal V}(-{\bf r}){\mathcal V}({\bf r}) &  = &  c({\bf r}) \ \mathbb{I}_{2} \, ,
\label{eq:inversion}\\
\tilde{\mathcal V}(-{\bf k})\tilde{\mathcal V}({\bf k}) &  = &  d({\bf k}) \ \mathbb{I}_{2}
\end{eqnarray}
\end{subequations}
where $c({\bf r}) = {\mathcal V}_{0}^{2}({\bf r})+{\mathcal C}_{x}^{2}({\bf r})+{\mathcal C}_{y}^{2}({\bf r})+{\mathcal C}_{z}^{2}({\bf r})$ and
$d({\bf k}) = \tilde{\mathcal V}_{0}^{2}({\bf k})+\tilde{\mathcal C}_{x}^{2}({\bf k})+\tilde{\mathcal C}_{y}^{2}({\bf k})+\tilde{\mathcal C}_{z}^{2}({\bf k})$ are scalar quantities. Furthermore, we notice that
\begin{subequations} 
\begin{eqnarray}
\label{eq:vomrk}
{\mathcal V}^{\dagger}({\bf r})  &=& {\mathcal V}(-{\bf r})
\, , 
\label{eq:vomrka} \\
\tilde{\mathcal V}^{\dagger}({\bf k})  &=&  \tilde{\mathcal V}({\bf k})\, .
\label{eq:vomrkb}
\end{eqnarray}
\end{subequations}
Since, due to Eqs.~(\ref{eq:inversion},\ref{eq:vomrka}), $c^{-1/2}({\bf r}){\mathcal V}({\bf r})$ is a unitary matrix, ${\mathcal V}({\bf r})$ belongs to the class of generalized unitary matrices, whereas $\tilde{\mathcal V}({\bf k})$ is a Hermitian matrix.

\section{Probability of quantum diffusion}
\label{sec:pqd}

The central quantity -- the intensity propagator $\Phi$ -- which we will use in this paper to characterize charge and spin diffusion is defined as follows
\begin{equation}
\Phi^{\sigma_1'\sigma_2',\sigma_1\sigma_2}(\varepsilon,\omega,{\bf r})  = 
 \overline{\sum_{m'}
g_{m',m}^{\sigma_1',\sigma_1(+)}(\varepsilon_1)g_{m,m'}^{\sigma_2,\sigma_2'(-)}(\varepsilon_2)
\delta\left({\bf r}-{\bf R}_{m' m} \right)}\label{eq:Phidef}
\end{equation}
in terms of the retarded (advanced) one-particle Green function
\begin{equation}
g_{m',m}^{\sigma',\sigma({\pm})}(\varepsilon) = 
\langle m' \sigma' \left|\frac{1}{\z_{\pm}-{\cal H}} \right| m \sigma \rangle \, .
\label{eq:smallg}
\end{equation}
We note $\varepsilon_{1,2}=\varepsilon \pm \hbar \omega/2$,  $\z_{\pm}=\varepsilon \pm i\eta$, and $\eta$ an infinitesimal positive quantity. The product of two (one-particle) Green functions appearing in the definition \eqref{eq:Phidef} warrants the denomination of two-particle Green function for the intensity propagator $\Phi$ (also called particle-hole Green function and particle-hole vertex function \cite{vol-woe,rev-vol-woe,akk-mont}).
The over-line in Eq.~(\ref{eq:Phidef}) stands for the average over the impurity configurations, assuming the position of the $N$ impurities to be random variables uniformly distributed on the volume $\Omega$. 
Due to the translational invariance obtained after impurity average, the propagator $\Phi$ is independent of the choice of the initial site $m$ in Eq.~(\ref{eq:Phidef}). 

From the intensity propagator $\Phi$, our physical quantities of interest can be extracted as follows: we consider as initial state a wave-packet
\begin{equation}
\label{eq:is}
|\psi_{\varepsilon,m,\sigma} \rangle = A \sum_{\nu} \ \langle \chi_{\nu} \ | \ m \sigma \rangle \ \exp{\left[-\frac{\left(\varepsilon_{\nu}-\varepsilon\right)^2}{4\sigma_{\varepsilon}^2}\right]} \ |\chi_{\nu}  \rangle \, ,
\end{equation}
describing an electron with energy $\varepsilon$ and spin $\sigma$ at site $m$,
where $\{|\chi_{\nu}\rangle\}$ is a complete basis of ${\cal H}$ with corresponding eigenenergies $\varepsilon_{\nu}$, $\sigma_{\varepsilon}$ is the energy-width of the wave-packet, and $A$ is a normalization constant. We are then interested in the impurity-averaged probability
\begin{equation}
P^{\sigma'\sigma}(\varepsilon,t,{\bf r})=\overline{\sum_{m'}\langle m'\sigma'|\varrho_t|m'\sigma'\rangle\delta\left({\bf r}-{\bf R}_{m' m} \right)}
\label{eq:Pphysical}
\end{equation}
to find, at a later time $t>0$, the electron with spin $\sigma'$  and at distance ${\bf r}$ from the initial  site, where the density operator 
$\varrho_t={\cal U}_t\varrho_0{\cal U}^\dagger$ denotes the state that results from the evolution ${\cal U}_t=\exp{[-i{\cal H}t/\hbar]}$ of the initial   density operator $\varrho_0=|\psi_{\varepsilon,m,\sigma} \rangle\langle\psi_{\varepsilon,m,\sigma}|$ at time $t=0$. 

The probability distribution governing the spatial (charge) diffusion is then given by 
\begin{equation}
\label{eq:spaceprob}
P^\sigma(\varepsilon,t,{\bf r}) = \sum_{\sigma'=\pm \sigma} P^{ \sigma',\sigma}(\varepsilon,t,{\bf r}) \, ,
\end{equation}
which, in general, depends on the direction of the initial spin $\sigma$. This dependence, however, vanishes in the limit of large distances $r$ (and large times $t$), where the spatial dynamics is described by an isotropic and spin-independent diffusion equation (as shown in Sec.~\ref{subsec:sdc} below). 

The spin probability is obtained from
\begin{equation}
P^{\sigma',\sigma}(\varepsilon,t) = \int {\rm d}{\bf r} \ P^{\sigma', \sigma}(\varepsilon,t,{\bf r}) \, .
\label{eq:spinprob}
\end{equation}
At large times $t$, the spin probability approaches its equilibrium value $1/2$ (exponentially in $t$), and the corresponding exponent defines the spin relaxation rate (see Sec.~\ref{subsec:sr}).

As shown in Appendix \ref{sec:AppendixA}, the probability $P^{\sigma'\sigma}(\varepsilon,t,{\bf r})$ is proportional to the Fourier transform of the intensity propagator 
$\Phi$:
\begin{equation}
P^{\sigma'\sigma}(\varepsilon,t,{\bf r})=  \frac{n_\mathrm{i}}{\rho(\varepsilon)} \ \frac{\hbar}{2\pi} \int_{-\infty}^{+\infty} {\rm d} \omega \ e^{-i\omega t} \
\Phi^{\sigma^{\prime}\sigma^{\prime},\sigma\sigma}(\varepsilon,\omega,{\bf r}) \, ,
\label{eq:defPrtssp}
\end{equation}
where $\rho(\varepsilon)$ denotes the impurity-averaged density of states. The latter, in turn, is obtained as the imaginary part
\begin{equation}
\label{eq:dos}
\rho(\varepsilon) = 
-\frac{n_\mathrm{i}}{\pi} \ {\rm Im}\left\{G^{\sigma,\sigma(+)}(\varepsilon)\right\} 
\end{equation}
of the local average Green function
\begin{equation}
G^{\sigma',\sigma (\pm)}(\varepsilon)=\overline{\left<m\sigma'\left|\frac{1}{\z_{\pm}-{\cal H}}\right|m\sigma\right>} \, .
\end{equation}
Due to spin symmetry, $G^{\sigma,\sigma}$ does not depend on $\sigma$, and this is why we have not attached a spin label to the density of states. 

From Eq.~(\ref{eq:defPrtssp}), it might seem that only diagonal elements of $\Phi$, i.e. $\sigma_1=\sigma_2$ and $\sigma_1'=\sigma_2'$ in Eq.~(\ref{eq:Phidef}), are relevant. 
However, in order to set up a self-consistent equation for $\Phi$ (see Eq.~(\ref{eq:bs00}) below), it is necessary to consider intermediate spin states which are not eigenstates of $\sigma_z$. Furthermore, the definition of $\Phi$ with four different indices makes it possible to generalize 
Eq.~(\ref{eq:spinprob}) to arbitrary initial and final spin states:
\begin{eqnarray}
R^{\sigma_{1}^{\prime}\sigma_{2}^{\prime},\sigma_{1}\sigma_{2}}(\varepsilon,t) & = & \frac{\hbar n_\mathrm{i}}{2\pi \rho(\varepsilon)} 
\int {\rm d}{\bf r}  
\int_{-\infty}^{+\infty} {\rm d} \omega
\nonumber \\
&  &
\ \times \ e^{-i\omega t} \ \Phi^{\sigma_{1}^{\prime}\sigma_{2}^{\prime},\sigma_{1}\sigma_{2}}(\varepsilon,\omega,{\bf r}) \, ,
\label{eq:spindm}
\end{eqnarray}
yielding the $(\sigma_{1}^{\prime}\sigma_{2}^{\prime})$ matrix-element of $\overline{\varrho}_t$, the impurity-averaged density operator (reduced to the spin subspace) that results from the evolution on an initial spin state characterized by $(\sigma_{1}\sigma_{2})$.

In the next chapter we set the basis for the self-consistent calculation of $G$ and $\Phi$ needed to determine the charge and spin diffusion.

\section{Locator expansion}
\label{sec:le}

A perturbative approach for $G$, and then for $\Phi$, can be addressed from the locator expansion 
\begin{equation}
\frac{1}{\z-{\cal H}}=\frac{1}{\z}+\frac{1}{\z}{\cal H}\frac{1}{\z}+\frac{1}{\z}{\cal H}\frac{1}{\z}{\cal H}\frac{1}{\z}+\dots \, ,
\label{eq:serieszH} 
\end{equation}
having the same form as in the spinless case \cite{mat-toy}, but now with   matrices having twice the dimension of the spinless ones. The uncoupled local Green functions are obviously scalar: $G_{00}^{\sigma',\sigma(\pm)}(\varepsilon)=G_{00}^{(\pm)}(\varepsilon) \ \delta_{\sigma',\sigma}$, with $G_{00}^{(\pm)}(\varepsilon)=1/\z_{\pm}$.

Since the operator ${\cal H}$ mediates transitions between different impurity sites (see Eq.~(\ref{eq:coupling})) each term of the series (\ref{eq:serieszH}) can be represented as the sum over paths connecting the initial site to the final site through an arbitrary sequence of intermediate sites with a given number of jumps, each of them associated with a hopping amplitude ${\mathcal V}$. Various diagrammatic prescriptions have been devised in the spinless case to graphically represent different terms of the locator expansion \cite{mat-toy,yonezawa64}. 
When averaging over the impurity positions, it is important to keep track of those sites which are visited more than once within a given path, and different conventions have been proposed to take into account this crucial subtlety. 

\begin{figure}
\begin{center}
\includegraphics[width=4cm]{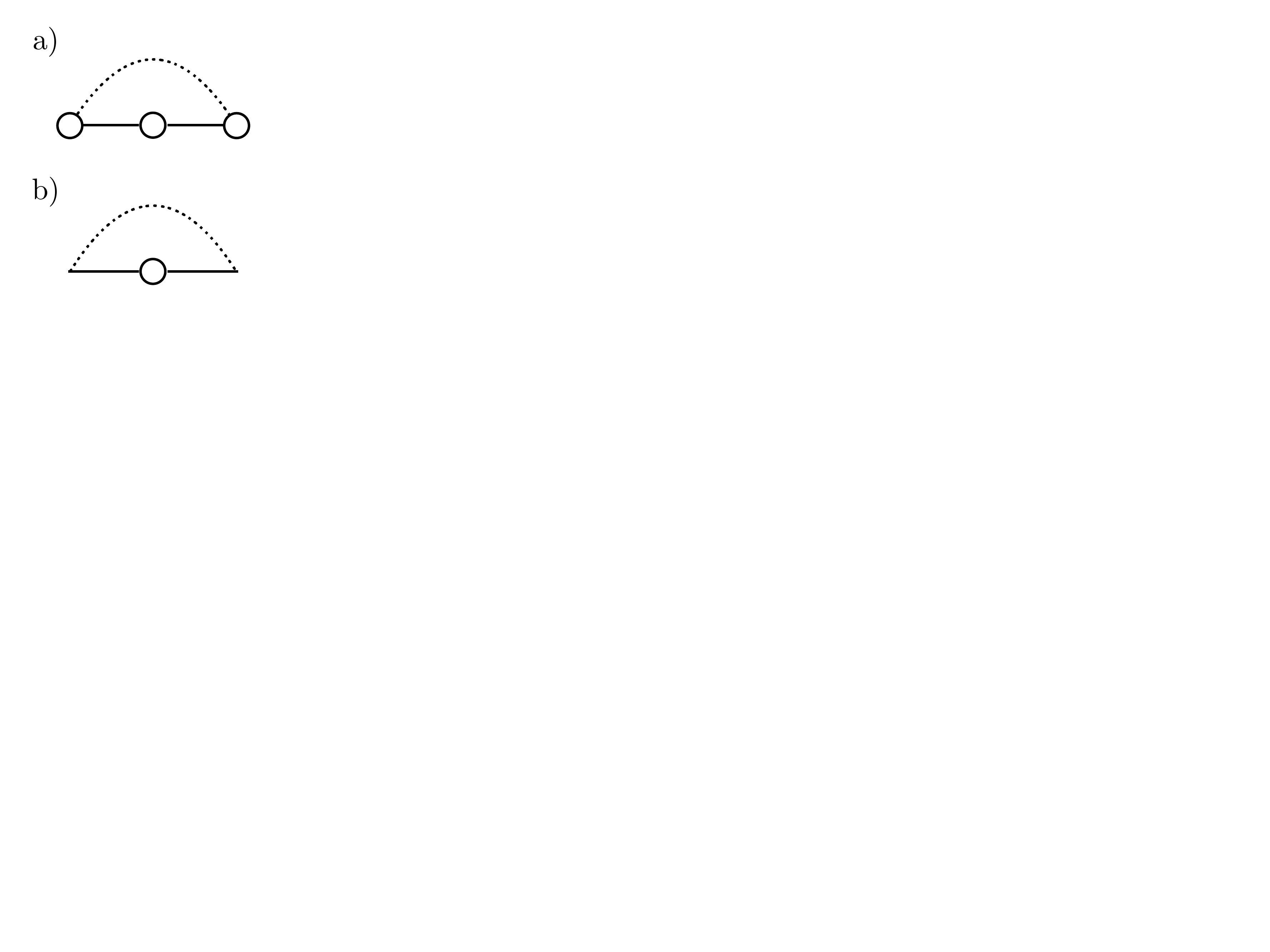}
\caption{(a) Example of an irreducible diagram of second order in the hopping amplitude contributing to the average local Green function $G^{(\pm)}(\varepsilon)$. The solid lines represent the hopping amplitude matrix ${\mathcal V}$, the circles stand for  $G_{00}^{(\pm)}(\varepsilon)=1/\z_{\pm}$, and the dotted lines indicate identical sites. (b) Self-energy $\Sigma^{(\pm)}(\varepsilon)$ corresponding to the irreducible
diagram for $G^{(\pm)}(\varepsilon)$ shown in (a).
}
\label{fig:fig1_GF}
\end{center}
\end{figure}

A simple way of systematizing the correspondence between the terms of the perturbation expansion and diagrams is to represent $G^{(\pm)}_{00}(\varepsilon)$ by a circle, 
${\mathcal V}^{\sigma'\sigma}({\bf R}_{m'm})$ by a solid horizontal line, and a dotted line to connect identical sites (i.e. sites which are visited more than once by a given path). As an example, the term
\begin{equation}
G^{\sigma'\!,\sigma(\pm)}(\varepsilon)=\overline{\sum_{m''\neq m}\sum_{\sigma''}\frac{1}{\z_{\pm}}\langle m\sigma'|{\mathcal V}|m''\sigma''\rangle \frac{1}{\z_{\pm}} \langle m''\sigma''|{\mathcal V}|m\sigma\rangle \frac{1}{\z_{\pm}}} \label{eq:Gexample}
\end{equation}
corresponding to the second-order contribution to the average local retarded (advanced) Green function $G^{\sigma',\sigma(\pm)}(\varepsilon)$ can be represented by the diagram shown in Fig.~\ref{fig:fig1_GF}(a), with the dotted line indicating that the initial and final impurity sites are the same. 

The diagram of Fig.~\ref{fig:fig1_GF}(a) has the property of being   'irreducible', since it cannot be decomposed into simpler (lower-order) ones by 'cutting' it at an intermediate Green function $G_{00}^{(\pm)}(\varepsilon)$. Any diagram contributing to the local average Green function can be factorized into its irreducible components by applying the 'cutting' recipe. Examples of 'reducible' diagrams are presented in Appendix \ref{apprid}. 

The sum of all irreducible diagrams defines the self-energy $\Sigma^{(\pm)}(\varepsilon)$, which is related to the average Green function through the Dyson equation:
\begin{equation}
G^{(\pm)}(\varepsilon)=\frac{1}{\z_{\pm}}+\frac{1}{\z_{\pm}}\Sigma^{(\pm)}(\varepsilon) \ G^{(\pm)}(\varepsilon) \, ,
\label{eq:sigmadef0}
\end{equation}
which can be rewritten as 
\begin{equation}
G^{(\pm)}(\varepsilon)=\frac{1}{\z_{\pm}-\Sigma^{(\pm)}(\varepsilon)}
\label{eq:sigmadef}
\end{equation}
Thus, each level of approximation chosen for $\Sigma^{(\pm)}(\varepsilon)$ generates the corresponding approximation for $G^{(\pm)}(\varepsilon)$, where arbitrary high orders in the hopping amplitude are included. Reciprocally, each irreducible diagram in the expansion of $G^{(\pm)}(\varepsilon)$ generates the corresponding contribution to $\Sigma^{(\pm)}(\varepsilon)$ by simply removing the two extreme (identical) circles. For instance, diagram Fig.~\ref{fig:fig1_GF}(b) is obtained from Fig.~\ref{fig:fig1_GF}(a) through the previous recipe. Thus, performing the impurity average in Eq.~\eqref{eq:Gexample}, we obtain the corresponding self-energy contribution
\begin{equation}
\Sigma^{(\pm)}(\varepsilon) = \frac{n_\mathrm{i}}{\z_{\pm}} 
\int {\rm d}{\bf r} \ {\mathcal V}(-{\bf r}){\mathcal V}({\bf r}) = \frac{n_\mathrm{i}}{\z_{\pm}} \int \frac{{\rm d}{\bf k}}{(2\pi)^3} {\mathcal V}^{2}({\bf k}) \, .
\label{eq:Sigma_example}
\end{equation}
From \eqref{eq:inversion}, it follows that $\Sigma^{(\pm)}(\varepsilon)$ is proportional to the identity matrix. Such a property is not restricted to the particular approximation of Eq.~\eqref{eq:Sigma_example}, but it is a general symmetry requirement of the local average Green function and the self-energy. While, in principle, Eqs.~(\ref{eq:sigmadef0},\ref{eq:sigmadef}) should be read as matrix equations for the $2\times 2$ matrices  $G^{(\pm)}(\varepsilon)$ and $\Sigma^{(\pm)}(\varepsilon)$, the symmetries (\ref{eq:symmetry2}) make that both matrices commute with the three Pauli matrices $\sigma_x$, $\sigma_y$, and $\sigma_z$. Therefore, they must be scalar quantities (proportional to the identity matrix), 
i.e. $G^{\sigma',\sigma(\pm)}(\varepsilon)=G^{(\pm)}(\varepsilon) \ \delta_{\sigma',\sigma}$. Herewith, our notation will not distinguish between the scalar quantity $G^{(\pm)}(\varepsilon)$ and the corresponding $2\!\times\!2$-matrix, obtained from $G^{(\pm)}(\varepsilon)$ by multiplication with $\mathbb{I}_{2}$. Likewise for $\Sigma^{(\pm)}(\varepsilon)$. 

\section{Self-consistent approach}
\label{sec:sca}

\begin{figure}
\begin{center}
\includegraphics[width=8.5cm]{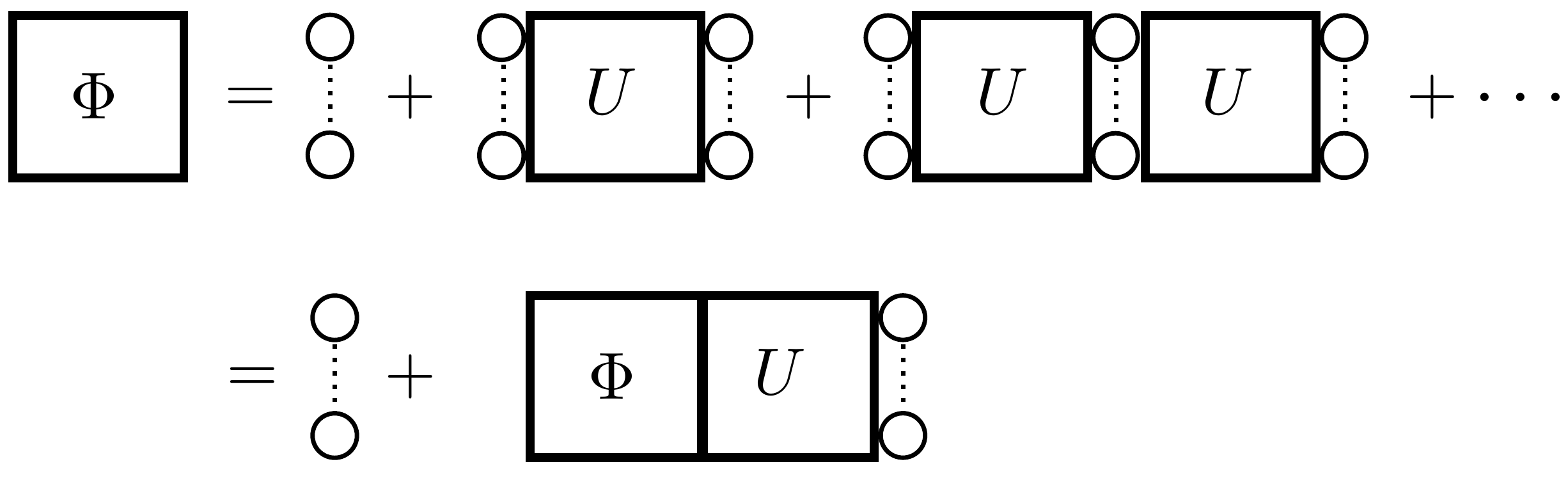}
\caption{Diagrammatic expression of the expansion of the intensity propagator $\Phi$ in terms of its irreducible component $U$ and the local average Green functions $G^{(\pm)}$ (first line), together with the resulting Bethe-Salpeter equation \eqref{eq:bs00} (second line).
}
\label{fig:phi}
\end{center}
\end{figure}

\subsection{Irreducible components of the one and two-particle Green functions}
\label{subsec:icotpgf}

A self-consistent approach for the one and two-particle Green functions (and their irreducible components) allows us to consider a restricted set of diagrams while respecting important constraints, like particle conservation. A self-consistent approximation to the local averaged Green function $G^{(\pm)}(\varepsilon)$ can be obtained if Eq.~\eqref{eq:sigmadef} is combined with an estimation of the local self-energy $\Sigma^{(\pm)}(\varepsilon)$ based on keeping a few simple diagrams, like the ones above discussed, but where the dots are now interpreted not as $G_{00}^{(\pm)}(\varepsilon)$, but as $G^{(\pm)}(\varepsilon)$. For instance, the diagram of Fig.~\ref{fig:fig1_GF}b for $\Sigma^{(\pm)}(\varepsilon)$ is extremely simple, but in its self-consistent form it effectively contains an infinite hierarchy of paths starting and ending at the same site (as it contains an intermediate $G^{(\pm)}(\varepsilon)$).

The previously introduced notion of irreducible diagram directly carries into the self-consistent approach by simply applying the 'cutting' criterion to the intermediate average local Green functions $G^{(\pm)}(\varepsilon)$. 
In addition, within the self-consistent approach, additional requirements appear in  order to avoid the double-counting of certain contributions; any self-consistent diagram kept for representing $\Sigma^{(\pm)}(\varepsilon)$ must have the property that it does not separate into unconnected parts whenever it is cut at {\em two} intermediate Green functions associated to the {\em same} impurity. This property -- which we call \lq two-point irreducible\lq\ -- 
allows us to build diagrams for the irreducible component $U$ of the intensity propagator $\Phi$. In Appendix \ref{apprid} we present examples of self-energy diagrams that do not satisfy the two-point irreducible requirement and we discuss the notion of 
irreducibility 
for the diagrams contributing to the intensity propagator.

The intensity propagator can be expressed as in Fig.~\ref{fig:phi} through the iteration of its irreducible component $U$ and Green functions. The latter must be understood as the self-consistent ones $G^{(\pm)}(\varepsilon)$ and concern the same impurity when related by a vertical dotted line. The connection between one and two-particle Green functions, and between their irreducible components, is an important aspect of the self-consistent approach. The recipe developed in Appendix \ref{apprbicip} establishes that the diagrams contributing to $U$ are constructed from those of $\Sigma^{(\pm)}(\varepsilon)$ by removing a single Green function (circle) from the latter, and then 'folding' all parts {\it left} from the removed Green function into the lower line.

\subsection{Bethe-Salpeter equation}
\label{subsec:bse}

The iteration represented in the first line of Fig.~\ref{fig:phi} can be written as a Bethe-Salpeter equation for the intensity propagator (second line in Fig.~\ref{fig:phi}):

\begin{widetext}
\begin{equation}
\label{eq:bs00}
\Phi^{\sigma_{1}^{\prime}\sigma_{2}^{\prime},\sigma_{1}\sigma_{2}}
(\varepsilon,\omega,{\bf r}) = 
G^{(+)}\left(\varepsilon_1\right) \ G^{(-)}\left(\varepsilon_2\right) \left[\delta_{\sigma_{1}^{\prime},\sigma_{1}}
\delta_{\sigma_{2}^{\prime}\sigma_{2}}\delta({\bf r})+
\sum_{\sigma_{1}^{\prime\prime}\sigma_{2}^{\prime\prime}}
\int{\rm d}{\bf r}'' \
\Phi^{\sigma_{1}^{\prime}\sigma_{2}^{\prime},
\sigma_{1}^{\prime\prime}\sigma_{2}^{\prime\prime}}
(\varepsilon,\omega,{\bf r}'') \ U^{\sigma_{1}^{\prime\prime}\sigma_{2}^{\prime\prime},\sigma_{1}\sigma_{2}}(\varepsilon,\omega,{\bf r}-{\bf r}'')\right]
 \, .
\end{equation} 
\end{widetext}
This simple form of the Bethe-Salpeter equation is valid for the case of local initial and final states (i.e. states which are localized on a single site) and a local self-energy. A more general form of the Bethe-Salpeter equation, which would be able to describe effects of spatial coherences (i.e. initial or final states which are not localized on a single site), would involve $\Phi$'s and $U$'s which depend on three position arguments instead of one. In this more general framework, it is also possible to consider non-local self-energies and Green functions, which, however, we do not develop in the present paper.

Switching to Fourier space, and using a $4\times 4$-matrix form, the Bethe-Salpeter equation becomes
\begin{equation}
\tilde{\Phi}(\varepsilon,\omega,{\bf q}) =  G^{(+)}\left(\varepsilon_1\right) \ G^{(-)}\left(\varepsilon_2\right) \left[1+\tilde{\Phi}(\varepsilon,\omega,{\bf q}) \ \tilde{U}(\varepsilon,\omega,{\bf q})\right] \, ,\label{eq:bs}
\end{equation}
and its formal solution can be written as
\begin{equation}
\tilde{\Phi}(\varepsilon,\omega,{\bf q})=\frac{1}{\left[G^{(+)}\left(\varepsilon_1\right) \ G^{(-)}\left(\varepsilon_2\right)
\right]^{-1}-\tilde{U}(\varepsilon,\omega,{\bf q})} \, .
\label{eq:Phisol}
\end{equation}

Solving the Bethe-Salpeter equation requires some kind of approximation for $\tilde{U}$. On the one hand, in the case of the spin probability, given by Eq.~\eqref{eq:spinprob}, the spatial integration implies that only the ${\bf q}\!=\!0$ values are relevant, and therefore we should determine $\tilde{\Phi}(\varepsilon,\omega,{\bf q}\!=\!0)$. On the other hand, the probability distribution governing the spatial diffusion, given by Eq.~\eqref{eq:spaceprob}, requires the knowledge of the small ${\bf q}$-values of $\tilde{\Phi}(\varepsilon,\omega,{\bf q})$.  

\subsection{Spin dynamics}
\label{subsec:sd}

Restricting ourselves to the case of ${\bf q}\!=\!0$, the self-consistent approach provides the connection between the irreducible components of the one and two-particle Green functions through the Ward identity

\begin{widetext}
\begin{equation}
\Sigma^{\sigma_3,\sigma_4(+)} \left(\varepsilon_1\right) -
\Sigma^{\sigma_3,\sigma_4(-)} \left(\varepsilon_2\right)
=\sum_{\sigma_1\sigma_2}
\left[G^{\sigma_1,\sigma_2(+)}\left(\varepsilon_1\right)-
G^{\sigma_1,\sigma_2(-)}\left(\varepsilon_2\right)
\right]\tilde{U}^{\sigma_1\sigma_2,\sigma_3\sigma_4}(\varepsilon,\omega,0) \, .
\label{eq:ward0}
\end{equation}
This identity (proven in Appendix \ref{appWI}) ensures, as shown below, the conservation of probability. Furthermore, for ${\bf q}=0$, the $4\times 4$ matrices $U$ and $\tilde{U}$ must remain invariant under the simultaneous rotation of an angle of $\pi$ around each of the three Cartesian axes, as well as under permutation of the axes labels. Therefore,
\begin{equation}
\label{eq:symmetry3}
\tilde{U}(\varepsilon,\omega,0) = 
\left(D_{\mu}^{(1/2)}(\pi) \otimes D_{\mu}^{(1/2)}(\pi)\right)\ \tilde{U}(\varepsilon,\omega,0) \ \left(D_{\mu}^{(1/2)}(\pi) \otimes D_{\mu}^{(1/2)}(\pi)\right)^{\dagger} = \left(P \otimes P^*\right) \ \tilde{U}(\varepsilon,\omega,0) \ \left(P \otimes P^*\right)^{\dagger} 
 \, ,
\end{equation}
with $\mu=(x,y,z)$ and symbol \lq$\otimes$\rq\ denoting the tensor product of $2\times 2$ matrices, i.e. $(A\otimes B)^{\sigma_1\sigma_2,\sigma_3\sigma_4}=A^{\sigma_1\sigma_3}B^{\sigma_2\sigma_4}$. The symmetries (\ref{eq:symmetry2}-\ref{eq:symmv4}) of the hopping amplitude matrix $\cal V$, from which the diagrammatic expansion for $\tilde{U}$ is constructed, determine the above-stated transformation properties of the latter. Solving the linear system of equations \eqref{eq:symmetry3} for the matrix elements of $\tilde{U}(\varepsilon,\omega,0)$, we obtain the general form
\begin{eqnarray}
\tilde{U}(\varepsilon,\omega,0) =  
\left(\begin{array}{cccc} \tilde{u}_1(\varepsilon,\omega) & 0 & 0 & \tilde{u}_2(\varepsilon,\omega)\\ 0 & \tilde{u}_1(\varepsilon,\omega)-\tilde{u}_2(\varepsilon,\omega) & 0 & 0\\ 0 & 0 & \tilde{u}_1(\varepsilon,\omega)-\tilde{u}_2(\varepsilon,\omega) & 0 \\ \tilde{u}_2(\varepsilon,\omega) & 0 & 0 & \tilde{u}_1(\varepsilon,\omega)\end{array}\right) \, .
\label{eq:Atilde}
\end{eqnarray}
\end{widetext}
Here, the matrix elements of $\tilde{U}$ are taken in the basis defined by the four basis vectors $(1,0,0,0)$, $(0,1,0,0)$, $(0,0,1,0)$, $(0,0,0,1)$ corresponding, respectively, to $(\sigma_1\sigma_2)=(++)$, $(+-)$, $(-+)$ and $(--)$. 

Inserting 
the form \eqref{eq:Atilde} of the matrix $\tilde{U}$
into Eq.~(\ref{eq:ward0}), and taking into account that $\Sigma$ and $G$ are both scalar quantities, we see that the symmetric form of $\tilde{U}$ is consistent with the Ward identity, provided that
\begin{equation}
\tilde{u}_1(\varepsilon,\omega)+\tilde{u}_2(\varepsilon,\omega)=\frac{
\Sigma^{(+)} \left(\varepsilon_1\right) -
\Sigma^{(-)} \left(\varepsilon_2\right)}{G^{(+)}\left(\varepsilon_1\right)-
G^{(-)}\left(\varepsilon_2\right)}\label{eq:wardsymm}
\end{equation}

From Eq.~(\ref{eq:Atilde}), we see that $\tilde{U}$ separates into well-defined blocks, and according to Eq.~\eqref{eq:Phisol}, $\tilde{\Phi}$ inherits this property. One of these blocks, spanned by the basis vectors $(1,0,0,0)$ and $(0,0,0,1)$, represents the subspace of diagonal reduced (spin) density operators $\varrho^{(\rm d)}=\alpha|+\rangle\langle+|+\beta|-\rangle\langle-|$, with real $\alpha$ and $\beta$ verifying $\alpha+\beta=1$. Therefore, if the initial spin density operator is diagonal, then also the final one has this property. We remark that the two only pure states of this subspace correspond to $(\alpha,\beta)=(1,0)$ or $(0,1)$, and they are able to evolve into mixed states because the quantum evolution after impurity average is no longer unitary. This $2\times 2$ block describes the spin lifetimes. The other two blocks of $\tilde{U}$, corresponding to the one-dimensional subspaces defined by the basis vectors $(0,1,0,0)$ and $(0,0,1,0)$, describe the evolution of spin coherences.

\subsection{Spin-relaxation rate}
\label{subsec:sr}

Restricting ourselves to the two-dimensional subspace of diagonal spin density operators, the eigenvalues of $\tilde{U}(\varepsilon,\omega,0)$ are 
\begin{equation}
\tilde{u}_{\pm}(\varepsilon,\omega)=\tilde{u}_1(\varepsilon,\omega) \pm \tilde{u}_2(\varepsilon,\omega)
\end{equation}
 corresponding, respectively, to the normalized eigenvectors ${\bf v}_{\pm} = (1/\!\sqrt{2})(1,0,0,\pm 1)$. Since $\tilde{\Phi}$ has the same eigenvectors as $\tilde{U}$, for $q=0$ the eigenvalues corresponding to the restricted subspace are 
\begin{equation}
\tilde{\phi}_\pm(\varepsilon,\omega)=\frac{1}{\left[G^{(+)}\left(\varepsilon_1\right) \ G^{(-)}\left(\varepsilon_2\right)\right]^{-1}-\tilde{u}_\pm(\varepsilon,\omega)}  \, .
\end{equation}
Using the relation
\begin{equation}
G^{(+)}\left(\varepsilon_1\right) \ G^{(-)}\left(\varepsilon_2\right) =\frac{G^{(+)}\left(\varepsilon_1\right)-G^{(-)}\left(\varepsilon_2\right)}
{(\varepsilon_2-i\eta)-\Sigma^{(-)}\left(\varepsilon_2\right)-(\varepsilon_1+i\eta)+\Sigma^{(+)}\left(\varepsilon_1\right)}\label{eq:greensfactor} \, ,
\end{equation}
together with Eqs.~(\ref{eq:wardsymm}) and \eqref{eq:dos}, we have
\begin{equation}
\label{eq:Phi0}
\tilde{\phi}_+(\varepsilon,\omega)  =  \frac{G^{(-)}\left(\varepsilon_2\right)-G^{(+)}\left(\varepsilon_1\right)}{\varepsilon_1-\varepsilon_2 +2i\eta} = \frac{2\pi\rho(\varepsilon)}{\hbar n_\mathrm{i}} \ \frac{i}{\omega+2i\eta} \, .
\end{equation}  
where the second equation holds for small $\omega$.
Transforming to the time domain, we write 
\begin{equation}
\tilde{\phi}_{+}(\varepsilon,t)=\frac{\hbar}{2\pi} \int_{-\infty}^{\infty} {\rm d}\omega \ e^{-i\omega t} \ \tilde{\phi}_{+}(\varepsilon,\omega)  = \frac{\rho(\varepsilon)}{n_\mathrm{i}} \, . 
\label{eq:norm1}
\end{equation}
As shown below, this equation expresses the probability conservation.
The spin-relaxation dynamics is determined by the second eigenvector 
\begin{equation}
\tilde{\phi}_-(\varepsilon,\omega)  =  \frac{G^{(-)}\left(\varepsilon_2\right)-G^{(+)}\left(\varepsilon_1\right)}{\varepsilon_1-\varepsilon_2 +2i\eta +2\Bigl[G^{(-)}\left(\varepsilon_2\right)-G^{(+)}\left(\varepsilon_1\right)\Bigr] \ \tilde{u}_2(\varepsilon,\omega)}\label{eq:phiminus} \, ,
\end{equation}
which in the limit of small $\omega$ and weak spin-orbit coupling takes the form
\begin{equation}
\tilde{\phi}_-(\varepsilon,\omega)  =
 \frac{2\pi\rho(\varepsilon)}{\hbar n_\mathrm{i}}\left(\frac{1}{-i\omega+4\pi \rho(\varepsilon) \tilde{u}_2(\varepsilon,0)/n_{\mathrm i}}\right) \, . 
\end{equation}

Transforming to the time domain, we have 
\begin{equation}
\tilde{\phi}_{-}(\varepsilon,t) = \frac{\rho(\varepsilon)}{n_\mathrm{i}} \ \exp{\left[-\frac{t}{\tau_{\mathrm{s}}(\varepsilon)}\right]} \, ,
\label{eq:norm2}
\end{equation}
where the spin-relaxation rate is given by
\begin{equation}
\frac{1}{\tau_{\mathrm{s}}(\varepsilon)}=\frac{4\pi\rho(\varepsilon)}{ \hbar n_\mathrm{i}} \ \tilde{u}_2(\varepsilon,0)
\label{eq:taus}
\end{equation}
For instance, an initial spin $\sigma\!=\!+$ corresponds to the basis vector $(1,0,0,0)=(1/\!\sqrt{2})({\bf v}_{+}\!+\!{\bf v}_{-})$. Therefore,  according to Eqs.~\eqref{eq:defPrtssp} and \eqref{eq:spinprob}, in the limit of large $t$ the spin probabilities write
\begin{equation}
P^{\pm,+}(\varepsilon,t) = \frac{1}{2} \left(1 \pm \exp{\left[-\frac{t}{\tau_{\mathrm{s}}(\varepsilon)}\right]}\right) \, ,
\label{eq:spinprobuu}
\end{equation}
where the eigenvalue \lq$+$\rq\ ensures the conservation of the total probability. 

\subsection{Spin coherences}
\label{subsec:sc}

As evident from the remaining elements in the central part of the $4 \times 4$-matrix $\tilde{U}$ (see Eq.~(\ref{eq:Atilde})),
the time evolution of the spin coherences is described by the same eigenvalue $\tilde{u}_{-}=\tilde{u}_1-\tilde{u}_2$. It then follows that the coherences are damped with the same rate $1/\tau_{\mathrm{s}}$ as the diagonal elements, as expected for a crystal with cubic symmetry. Therefore, according to Eqs.~\eqref{eq:spindm} and \eqref{eq:spinprobuu}, the time evolution of the impurity-averaged density operator (reduced to the spin subspace), at long times $t$ and fixed energy $\varepsilon$ is
\begin{equation}
\overline{\varrho}_{t}=\left(\begin{array}{cc}\frac{1}{2}+\left(\varrho^{++}_{t=0}-\frac{1}{2}\right)e^{-t/\tau_{\mathrm{s}}(\varepsilon)} & \varrho^{+-}_{t=0}e^{-t/\tau_{\mathrm{s}}(\varepsilon)} \\
 \varrho^{-+}_{t=0}e^{-t/\tau_{\mathrm{s}}(\varepsilon)} & \frac{1}{2}+\left(\varrho^{--}_{t=0}-\frac{1}{2}\right)e^{-t/\tau_{\mathrm{s}}(\varepsilon)}\end{array}\right) \, ,
 \end{equation}
where $\varrho^{\sigma_1 \sigma_2}_{t=0}$ are the matrix elements of the initial spin density operator.

\subsection{Spatial diffusion coefficient}
\label{subsec:sdc}

As discussed in Sec.~\ref{subsec:bse}, studying the spatial diffusion requires the small-${\bf q}$ expansion of the matrix equation \eqref{eq:Phisol}.
For this purpose, we expand $\tilde{U}(\varepsilon,\omega,{\bf q})$ up to second order in $q$
\begin{equation}
\tilde{U}(\varepsilon,\omega,{\bf q})\simeq \tilde{U}(\varepsilon,\omega,0)+\sum_{\mu} q_\mu \hat{u}_\mu(\varepsilon,\omega)+
\sum_{\mu\mu'} q_\mu q_{\mu'} \hat{u}_{\mu\mu'}(\varepsilon,\omega)
\end{equation}
where $\hat{u}_\mu(\varepsilon,\omega)=\partial_{q_\mu}\left.\tilde{U}(\varepsilon,\omega,{\bf q})\right|_{{\bf q}=0}$ and
 $\hat{u}_{\mu\mu'}(\varepsilon,\omega)=\frac{1}{2}\partial_{q_\mu}\partial_{q_{\mu'}}\left.\tilde{U}(\varepsilon,\omega,{\bf q})\right|_{{\bf q}=0}$.
 Here, $\mu$ and $\mu'$ stand for the three spatial directions $(x,y,z)$.
 The symmetries discussed in Sec.~\ref{subsec:sha} above imply relations for the $\hat{u}_\mu(\varepsilon,\omega)$'s and 
 $\hat{u}_{\mu\mu'}(\varepsilon,\omega)$'s similar to those of Eq.~(\ref{eq:symmetry3}), which, however, involve additional minus signs whenever the corresponding symmetry, see Eqs.~(\ref{eq:symmetry2}), involves an inversion of the spatial coordinate $\mu$ or $\mu'$ with respect to which a derivative is taken, e.g.
\begin{eqnarray}
\hat{u}_x(\varepsilon,\omega) & = & - 
\left(D_{y}^{(1/2)}(\pi) \otimes D_{y}^{(1/2)}(\pi)\right)\ \hat{u}_x(\varepsilon,\omega) \nonumber\\
& & \left(D_{y}^{(1/2)}(\pi) \otimes D_{y}^{(1/2)}(\pi)\right)^{\dagger} 
\, .
\end{eqnarray}
since a $c_2$ rotation around the $y$-axis inverts the sign of $x$.
Moreover, the cyclic permutation symmetry (see Eq.~(\ref{eq:symmv4})) now implies relations between  different spatial components, e.g.
\begin{equation}
\hat{u}_{xy}(\varepsilon,\omega) = \left(P \otimes P^*\right) \ \hat{u}_{yz}(\varepsilon,\omega) \ \left(P \otimes P^*\right)^{\dagger} \ .
\end{equation} 
Solving the linear sets of equations resulting from these symmetry relations for the matrix elements of 
the $\hat{u}_\mu(\varepsilon,\omega)$'s and 
 $\hat{u}_{\mu\mu'}(\varepsilon,\omega)$'s, we see
that the first order terms and the non-diagonal components of the second order terms vanish, i.e. $\hat{u}_\mu(\varepsilon,\omega)=0$ for $\mu=x,y,z$ and $\hat{u}_{\mu\mu'}(\varepsilon,\omega)=0$ for $\mu\neq \mu'$. Concerning the diagonal components of the 
 second order terms, it follows that the vector ${\bf v}_+$ is an eigenvector of $\hat{u}_{\mu\mu}(\varepsilon,\omega)$ with the same eigenvalue for $\mu=x,y$ and $z$, respectively. Due to Eq.~(\ref{eq:Phisol}),  ${\bf v}_+$  remains also eigenvector of the intensity propagator $\Phi$ for small $q$. The corresponding eigenvalue $\phi_+$ yields the partial trace over the spin degree of freedom, since, according to Eqs.~(\ref{eq:spaceprob},\ref{eq:defPrtssp}):
 \begin{equation}
 \tilde{\phi}_+(\varepsilon,\omega,{\bf q})  =  \frac{2\pi \rho(\varepsilon)}{\hbar n_{\mathrm i}}\int{\rm d}\omega\, e^{i\omega t}\int{\rm d}{\bf r}\, e^{i{\bf q}\cdot{\bf r}}\, P^\sigma(\varepsilon,{\bf r},t) \, ,
\end{equation}
provided that ${\bf v}_+$ is an eigenvector of $\tilde{\Phi}(\varepsilon,\omega,{\bf q})$.
For small $\omega$ and $q$, this eigenvalue is given by:
\begin{equation}
\label{eq:Phismallwq}
\tilde{\phi}_+(\varepsilon,\omega,{\bf q}) = \frac{2\pi\rho(\varepsilon)}{ \hbar n_{\mathrm i} }\left(\frac{1}{-i\omega+q^2 {\cal D}(\varepsilon)}\right)\ \, , 
\end{equation}
which has the form of a diffusion pole with diffusion constant
\begin{equation}
{\cal D}(\varepsilon)=-\frac{\pi\rho(\varepsilon)}{\hbar n_{\rm i}} \partial^2_{q_\mu}\left.\left(\tilde{U}^{++,++}(\varepsilon,0,{\bf q})+\tilde{U}^{++,--}(\varepsilon,0,{\bf q})\right)\right|_{{\bf q}=0}
\label{eq:Dgeneral}
\end{equation}
Due to the cubic symmetry of the zinc-blende structure, the diffusion constant is independent of the spatial direction $\mu$ 
and of the direction of the initial spin $\sigma$.

It is instructive to reformulate Eq.~(\ref{eq:Dgeneral}) in position space as
\begin{equation}
{\cal D}(\varepsilon)=\frac{\int {\rm d}{\bf r}~r^2 p(\varepsilon,{\bf r})}{6\tau(\varepsilon)} \, ,
\label{eq:diffusionr}
\end{equation}
with
\begin{equation}
p(\varepsilon,{\bf r})=\left|G^{(+)}(\varepsilon)\right|^2 \Bigl(U^{++,++}(\varepsilon,0,{\bf r})+U^{++,--}(\varepsilon,0,{\bf r})\Bigr) \, ,
\label{eq:steplength}
\end{equation}
and 
\begin{equation}
\tau(\varepsilon)=-\frac{\hbar}{2~{\rm Im}\left\{\Sigma^{(+)}(\varepsilon)\right\}}\ .
\label{eq:steptime}
\end{equation}
The Ward identity (\ref{eq:ward0}) guarantees the normalization of $p(\varepsilon,{\bf r})$, i.e. $\int{\rm d}{\bf r}~p(\varepsilon,{\bf r})=1$. If, in addition,
$p(\varepsilon,{\bf r})\geq 0$ for all ${\bf r}$, the quantity $p(\varepsilon,{\bf r})$ can be interpreted as a classical probability distribution. In this case, the diffusion constant (\ref{eq:diffusionr}) is that of a classical random walk characterized by a step-length distribution $p(\varepsilon,{\bf r})$ and a hopping time $\tau(\varepsilon)$. More precisely, $p(\varepsilon,{\bf r}'-{\bf r})$ defines the probability density for a single step of the random walk from ${\bf r}$ to ${\bf r}'$, whereas $\tau(\varepsilon)$ specifies the time between two successive steps. 
Eq.~(\ref{eq:diffusionr}) then reproduces the standard three-dimensional expression $\langle R^2\rangle=6 {\cal D}(\varepsilon)t$ for the mean squared displacement $\langle R^2\rangle = N \langle r^2\rangle$ after a number of hops $N=t/\tau(\varepsilon)$, where $\langle r^2\rangle=\int{\rm d}{\bf r}~r^2 p(\varepsilon,{\bf r})$ is the mean squared displacement of a single step.

\section{Simplest self-consistent approximation}
\label{sec:simple}

The general theory outlined in the previous chapter allows us to calculate the spin relaxation rate and the spatial diffusion constant from the self-energy $\Sigma$ and  the irreducible component $U$ of the intensity propagator. In order to determine $\Sigma$ and $U$, we first need to select certain diagrams. To start with, we will consider
a particularly simple choice of diagrams in order to illustrate our theory.

\begin{figure}
\includegraphics[width=8.5cm]{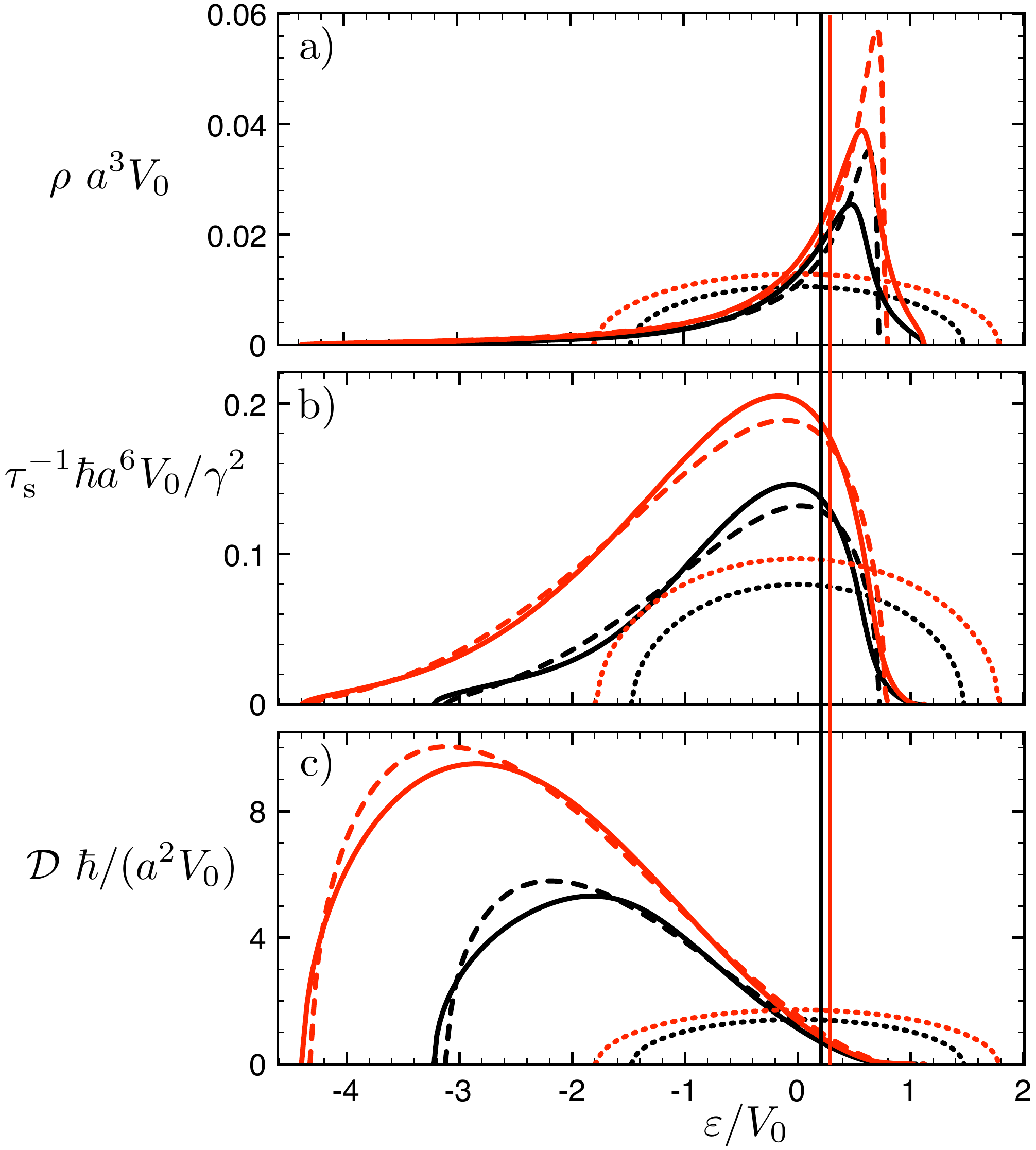}
\caption{(a) Density of electronic states $\rho$, (b) spin-relaxation rate $\tau_{\mathrm{s}}^{-1}$, (c) diffusion coefficient ${\cal D}$ in the impurity band as a function of energy for the three approaches developed in this work: the 
simplest self-consistent approximation (SSCA, dotted), 
the loop-corrected self-consistent approximation (LCSCA, dashed), and the repeated-scattering-corrected self-consistent approximation (RSCSCA, solid). Results are presented for two dimensionless impurity densities 
${\mathcal N}_\mathrm{i}=0.29^3$ (black) and $0.33^3$ (red). Energies are measured using the isolated impurity level as origin and are expressed in units of $V_0$ (twice the ionization energy). The other physical constants used to define the scales of the figure are the effective Bohr radius $a$ and the Dresselhaus coupling constant 
$\gamma\ll a^3 V_0$. 
The vertical lines mark the position of the Fermi energy corresponding to the two considered impurity densities for the RSCSCA, and are very close to those of the LCSCA. The Fermi energy for the 
SSCA
is that of the isolated impurity level $\varepsilon_{00}=0$.
}
\label{fig:rhoeps_and_taueps}
\end{figure}

\subsection{Self-energy and density of states}
\label{sec:sedos}

The 
simplest self-consistent approximation (SSCA) for $\Sigma^{(\pm)}(\varepsilon)$ is that of Fig.~\ref{fig:fig1_GF}(b), where the local averaged Green function $G^{(\pm)}(\varepsilon)$ is understood as being the self-consistent one. In this section we will show that this simple approximation reproduces the analytical result \eqref{eq:taus_vs_dens} for the spin relaxation rate.

The diagram of Fig.~\ref{fig:fig1_GF}(b), representing the processes where the electron hops from site $m$ to another site $m''\neq m$, and then back to $m$, translates into
\begin{equation}
\Sigma^{(\pm)}(\varepsilon)  =  \sum_{m''\sigma''} \overline{\langle m\sigma|{\mathcal V}|m''\sigma''\rangle \ G^{(\pm)}(\varepsilon) \ \langle m''\sigma''|{\mathcal V}|m\sigma\rangle} \, , \label{eq:sigma00}
\end{equation}
which is related with Eq.~\eqref{eq:Gexample} by the rule enunciated in Sec.~\ref{sec:le} for linking the self-energy and the Green function at each level of approximation. The central $G_{00}^{(\pm)}(\varepsilon)$ of Eq.~\eqref{eq:Gexample} is replaced by the self-consistent one $G^{(\pm)}(\varepsilon)$, which is diagonal in spin indices, and therefore the self-consistent version of Eq.~\eqref{eq:Sigma_example} writes 
\begin{equation}
\Sigma^{(\pm)}(\varepsilon)  =  
n_\mathrm{i}  \ G^{(\pm)}(\varepsilon)\int{\rm d}{\bf r}~{\mathcal V}(-{\bf r})\ {\mathcal V}({\bf r})  
= \alpha \ G^{(\pm)}(\varepsilon) \, , 
\label{eq:sigma00b}
\end{equation}
where, according to \eqref{eq:meh0}, \eqref{eq:Cxyz}, and \eqref{eq:inversion},
\begin{equation}
\alpha = n_\mathrm{i}  \int{\rm d}{\bf r} \ c({\bf r}) 
= \pi n_\mathrm{i} \left(7a^3V_0^2+\frac{2}{7a^3}\gamma^2\right) \, . \label{eq:alpha}
\end{equation}

Using Eqs.~\eqref{eq:sigmadef} and \eqref{eq:sigma00b}, we have
\begin{equation}
\Sigma^{(\pm)}(\varepsilon)=\frac{\alpha}{\z_{\pm}-\Sigma^{(\pm)}(\varepsilon)} \, .
\label{eq:sigma0}
\end{equation}
The solution of this self-consistent equation for the retarded self-energy $\Sigma^{(+)}(\varepsilon)$ with negative imaginary part reads
\begin{equation}
\Sigma^{(+)}(\varepsilon)=\frac{1}{2}\left(\z_+- i \sqrt{4\alpha-\z_+^2}\right) \, .
\label{eq:sigma0sol}
\end{equation}

The resulting density of states stemming from Eqs.~\eqref{eq:dos} and \eqref{eq:sigma0sol} is 
\begin{eqnarray}
\rho(\varepsilon) & = & \left\{\begin{array}{cll } n_\mathrm{i} \frac{\sqrt{4\alpha-\varepsilon^2}}{2\pi\alpha} & {\rm if} & \varepsilon^2<4\alpha \, , \\
0 & {\rm if} &\varepsilon^2\geq 4\alpha \, ,
\end{array} \right.
\label{eq:rho0}
\end{eqnarray}
and has the shape of a semicircle with radius $2\sqrt{\alpha}$ (dotted lines in Fig.~\ref{fig:rhoeps_and_taueps}(a). Notice that the density of states is normalized such that $\int_{-\infty}^\infty{\rm d}\varepsilon~\rho(\varepsilon)=n_\mathrm{i}$, corresponding to one electronic state per impurity site and spin species.

\begin{figure}
\begin{center}\includegraphics[width=5cm]{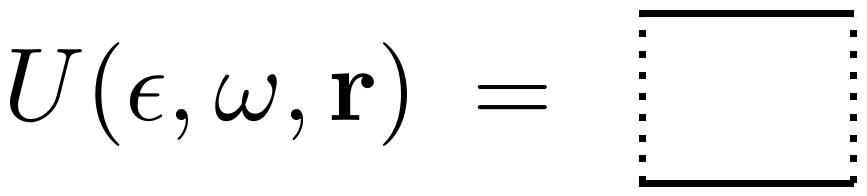}
\caption{Irreducible component of the intensity propagator within the 
SSCA, obtained by applying the 'cut and fold' procedure to the self-consistent self-energy of Fig.~\ref{fig:fig1_GF}(b). The solid horizontal upper (lower) line stands for the hopping amplitude matrix ${\mathcal V}({\bf r})$ (${\mathcal V}^{*}({\bf r})$), the dotted vertical lines indicate identical sites.
}
\label{fig:irrU}
\end{center}
\end{figure}

\subsection{Irreducible component of the intensity propagator}
\label{sec:icip}

The recipe for constructing the irreducible component $U$ of the intensity propagator from the corresponding diagram representing $\Sigma$ (see App. \ref{tascale}) yields the diagram of Fig.~\ref{fig:irrU}, which can be expressed as
\begin{equation}
U^{\sigma^{\prime}_1\sigma^{\prime}_2,\sigma_1\sigma_2}(\varepsilon,\omega,{\bf r})   =   
 n_\mathrm{i}  {\mathcal V}^{\sigma^{\prime}_1\sigma_1}({\bf r})
\left({\mathcal V}^{\sigma^{\prime}_2\sigma_2}({\bf r})\right)^* \, .
\label{eq:A00}
\end{equation}
That is, $U(\varepsilon,\omega,{\bf r})  =   n_\mathrm{i}{\mathcal V}({\bf r})\otimes{\mathcal V}^*({\bf r})$ and the matrix $U$ in Fourier space can be written as
\begin{equation}
\tilde{U}(\varepsilon,\omega,{\bf q})  = n_\mathrm{i}  \int\frac{{\rm d}{\bf k}} {(2\pi)^3} \ \tilde{\mathcal V}({\bf k}_+)\otimes \tilde{\mathcal V}^*({\bf k}_-) \, ,
\label{eq:A0}
\end{equation}
where ${\bf k}_\pm={\bf k}\pm{\bf q}/2$. 

\subsection{Spin relaxation rate}
\label{sec:srr}

\begin{center}
\begin{figure}
\includegraphics[width=8.5cm]{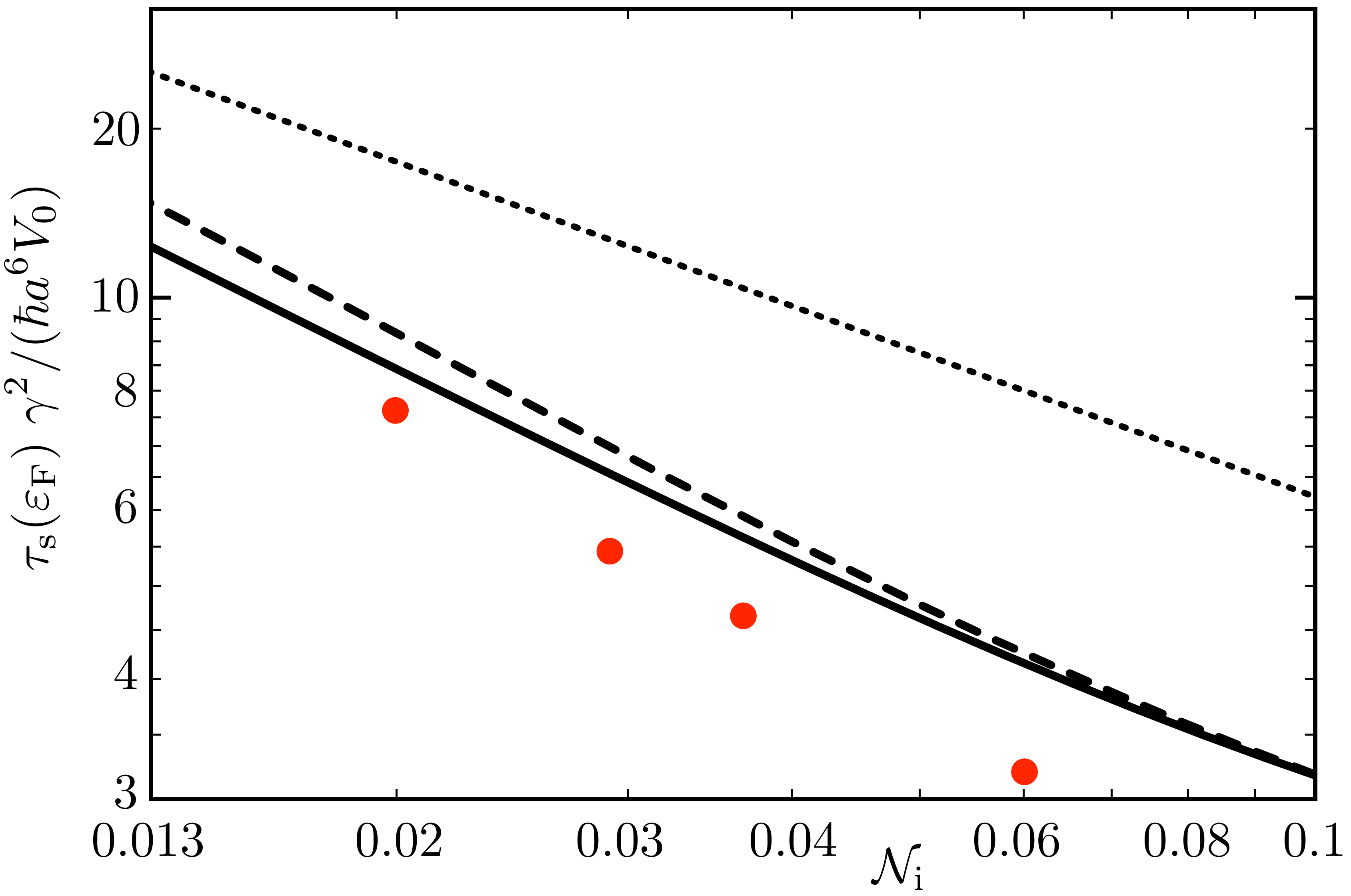}
\caption{Spin-relaxation time at the Fermi energy $\tau_{\mathrm{s}}(\varepsilon_{\mathrm F})$, for a half-filled impurity band, as a function of the dimensionless impurity density $\mathcal{N}_\mathrm{i}=n_\mathrm{i}a^3$ for the three approaches developed in this work: the 
SSCA (Eq.~\eqref{eq:taulosc}, dotted), the LCSCA (dashed), and the RSCSCA (solid). The red dots are the numerical results of Ref.~\onlinecite{prl2012}. The theory only applies to the interval $0.017 < \mathcal{N}_\mathrm{i} < 0.07$
(i.e. between the critical density ${\mathcal  N}_{\rm c}\simeq 0.017$ of the Mott transition and the hybridization density ${\mathcal N}_{\rm h}\simeq 0.07$); 
the lowest density results are presented to show the importance of the repeated-scattering correction in this regime. 
}
\label{fig:agreement}
\end{figure}
\end{center}

The spin dynamics is described by the case of ${\bf q}=0$, where the integral of Eq.~(\ref{eq:A0}), can be readily done. The resulting $\tilde{U}(\varepsilon,\omega,0)$ respects the form given in Eq.~(\ref{eq:Atilde}), with
\begin{subequations}
\begin{eqnarray}
\tilde{u}_1(\varepsilon,\omega) & = &  n_\mathrm{i} \int\frac{{\rm d}{\bf k}}{(2\pi)^3} \left[\tilde{\mathcal V}^2_0({\bf k})+\left|\tilde{\mathcal C}_z\right|^2({\bf k})\right] \nonumber \\
&  & \hspace{1cm} = 7\pi a^3 n_\mathrm{i} V_0^2+\frac{2\pi n_\mathrm{i}}{21a^3} \ \gamma^2 \label{eq:upm1}\\
\tilde{u}_2(\varepsilon,\omega) &=&  n_\mathrm{i} \int\frac{{\rm d}{\bf k}}{(2\pi)^3} \left[\left|\tilde{\mathcal C}_x\right|^2({\bf k})+\left|\tilde{\mathcal C}_y\right|^2({\bf k})\right]\nonumber \\
& & \hspace{1cm} =\frac{4\pi n_\mathrm{i}}{21a^3}\gamma^2 \label{eq:upm2}
\end{eqnarray}
\end{subequations}
Inserting Eq.~(\ref{eq:upm2}) into the general expression (\ref{eq:taus}), we obtain the energy-dependent spin relaxation rate $\tau_{\mathrm{s}}^{-1}(\varepsilon)$ (dotted lines in Fig.~\ref{fig:rhoeps_and_taueps}(b)), which follow a semi-circle law due to the proportionality with the density of states. For uncompensated semiconductors the impurity band is half-filled (the Fermi energy is $\varepsilon_{\mathrm F}=\varepsilon_{00}=0$), and assuming $\gamma\ll a^3 V_0$ we have
\begin{equation}
\frac{1}{\tau_{\mathrm{s}}(0)}=\frac{16}{21}\sqrt{\frac{\pi}{7}} \ \frac{\gamma^2}{a^6V_0\hbar} \ \mathcal{N}_\mathrm{i}^{1/2} 
\simeq 0.51 \ \frac{\gamma^2}{a^6V_0\hbar} \ \mathcal{N}_\mathrm{i}^{1/2} \, .
\label{eq:taulosc}
\end{equation}
Comparing Eqs.~\eqref{eq:taus_vs_dens} and \eqref{eq:taulosc}, we notice that the 
SSCA
for the spin relaxation rate reproduces the phenomenological result, up to a numerical factor. The difference between the prefactors of both equations is not surprising, since in the phenomenological approach some of the numerical constants are arbitrary. The $\mathcal{N}_\mathrm{i}$-dependence of the SSCA
spin-relaxation time $\tau_{\mathrm{s}}(0)$ is presented in Fig.~\ref{fig:agreement} (dotted line), together with the numerical results of Ref.~\onlinecite{prl2012} (red dots) and those of the approaches to be developed in the sequel. 

\subsection{Spatial diffusion}
\label{sec:sd}

The spatial diffusion coefficient is obtained from Eq.~(\ref{eq:diffusionr}) with step-length distribution
\begin{equation}
p(\varepsilon,{\bf r})=\frac{n_{\rm i}}{\alpha}\left({\mathcal V}_0^2({\bf r})+{\mathcal C}_x^2({\bf r})+{\mathcal C}_y^2({\bf r})+{\mathcal C}_z^2({\bf r})\right)
\end{equation}
and time
\begin{equation}
\tau(\varepsilon)=\frac{\hbar n_{\rm i}}{2\alpha\pi\rho(\varepsilon)}\ .
\end{equation}
Since $p(\varepsilon,{\bf r})\geq 0$ for all ${\bf r}$, the spatial diffusion dynamics described by the SSCA can be interpreted as a classical random walk.
Note, however, that  $p(\varepsilon,{\bf r})$ differs from the step-length distribution $\exp(-r/\ell_\varepsilon)/(4\pi r^2 \ell_\varepsilon)$ of a  random walk with mean free path $\ell_\varepsilon$, as described by the
classical Boltzmann equation. This is not surprising, since the Boltzmann equation applies to continuous systems whereas we are dealing with a discrete network of impurities.
Using the expressions \eqref{eq:fth} for $\tilde{\mathcal V}_0({\bf k})$ and $\tilde{\mathcal C}_{\mu}({\bf k})$, the diffusion constant results as
\begin{equation}
{\cal D}(\varepsilon) = \left(\frac{27a^5}{2}V_0^2+\frac{4}{3a}\gamma^2\right) \pi^2 \rho(\varepsilon) \, .
\end{equation}
In the simple scheme of the 
SSCA, the diffusion constant and the density of states can be analytically calculated, and we therefore obtain the spin-dependent correction to the spatial diffusion. Since $\gamma\ll V_0 a^3$, such a correction is very small. The diffusion coefficient (dotted lines in Fig.~\ref{fig:rhoeps_and_taueps}(c)), being proportional to the density of states, follows the semi-circle law of the latter.

\section{Loop-corrected self-consistent approximation}
\label{sec:lcsca}

\subsection{Self-energy and density of states}
\label{sec:sedos2}

\begin{figure}
\begin{center}
\includegraphics[width=8cm]{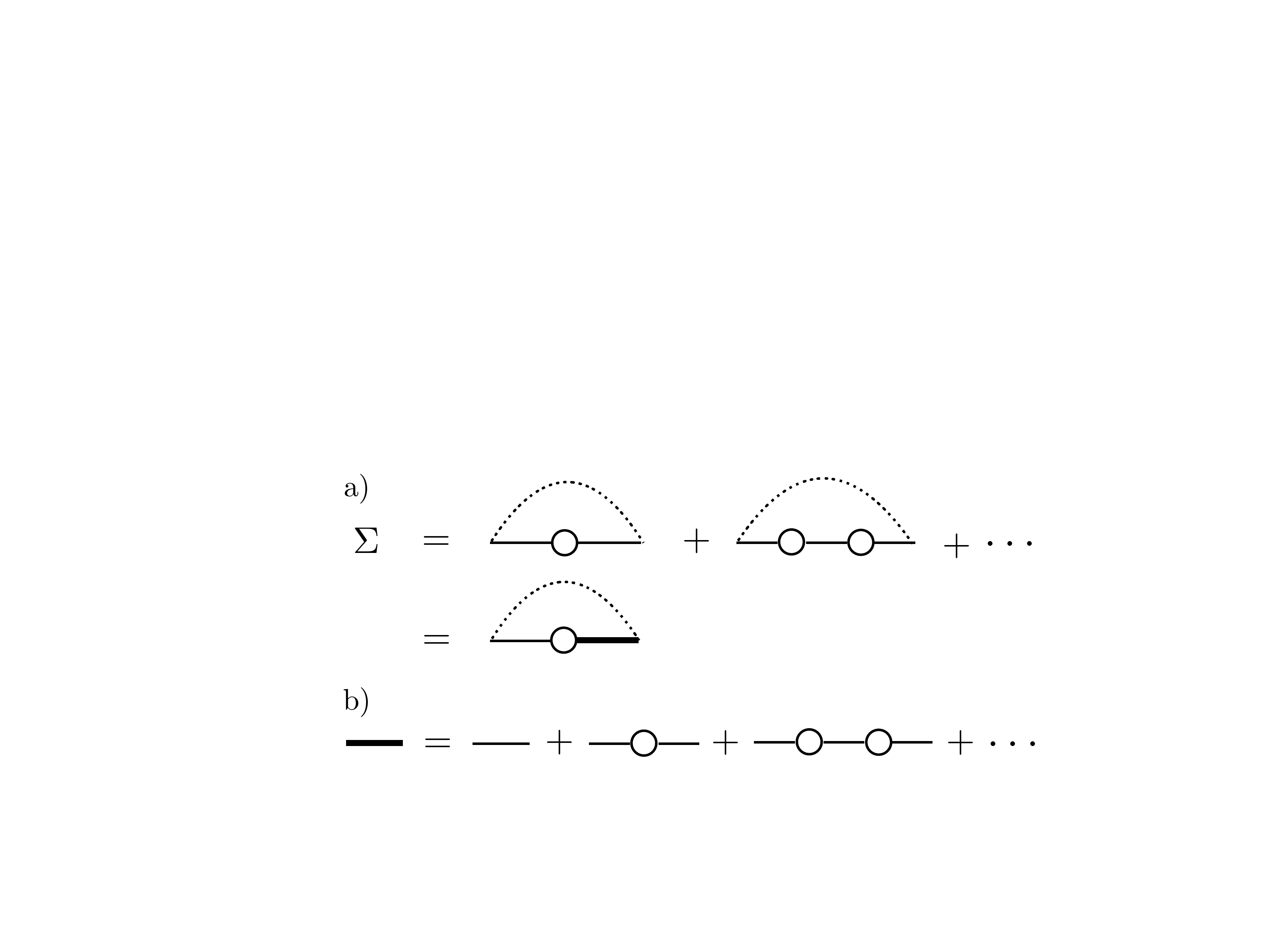}
\caption{a) Self-energy obtained by considering loops of arbitrary length representing processes where the electron visits different  impurities before hopping back to the starting site (LCSCA). b) Diagrammatic expression of the renormalized hopping amplitude (thick horizontal line) as an expansion of infinite order in the bare hopping amplitude (thin horizontal lines). 
}
\label{fig:Sigma_loops}
\end{center}
\end{figure}

As shown in the previous section, the diagram of Fig.~\ref{fig:fig1_GF}(b) for the self-energy provides analytical results for the density of states, the spin-relaxation time and the diffusion constant, which can be used to estimate the relevant orders of magnitude. These estimates, however, are not expected to yield quantitatively precise results, neither for low nor for large impurity densities. In this chapter, we will present another approximation which becomes exact in the limit of very large impurity densities. For large impurity densities, processes in which the electron visits more than one impurity before hopping back to the starting one, have to be considered. At the same time, the large number of impurities allows us to neglect repeated scatterings from the same impurity.
The one-loop approximation of Fig.~\ref{fig:fig1_GF}(b) is then extended to the sum of loops of arbitrary length, as presented in Fig.~\ref{fig:Sigma_loops}(a). As before, the dots represent the local averaged self-consistent Green functions and the dotted line connecting the extreme points identify the initial and final sites.    

The loop-corrected self-consistent approximation (LCSCA) of $\Sigma(\varepsilon)$ that we develop in this section for the interaction \eqref{eq:coupling} has been used, in the spinless case, by Matsubara and Toyozawa \cite{mat-toy}, in order to obtain the density of states in the impurity band, as well as the conductivity within the diffusion approximation. 
The MT scheme, ignoring cross 
diagrams, is referred to as a single-site approximation \cite{gas-cyr}.

The diagrams of Fig.~\ref{fig:Sigma_loops}(a) are, in the spinless case, equivalent to those introduced in Ref.~\onlinecite{mat-toy}. Moreover, they readily allow the matrix generalization to treat spin-dependent interactions and they are appropriate to use the self-consistent rules for building the irreducible components of the intensity propagator. 

Switching into Fourier space and applying the convolution theorem, the self-energy represented in Fig.~\ref{fig:Sigma_loops}(a) can be expressed as a geometrical series
\begin{equation}
\Sigma^{(\pm)}(\varepsilon) = \int\frac{{\rm d}{\bf k}}{(2\pi)^3} \tilde{\mathcal V}({\bf k}) \sum_{n=1}^\infty \left[n_\mathrm{i} \ 
G^{(\pm)}(\varepsilon) \ \tilde{\mathcal V}({\bf k})\right]^n \, .
\label{eq:Sigmadsca}
\end{equation}
Using Eq.~\eqref{eq:sigmadef}, we obtain the self-consistent condition for the self-energy
\begin{equation}
\Sigma^{(\pm)}(\varepsilon) =  n_\mathrm{i} \int\frac{{\rm d}{\bf k}}{(2\pi)^3} \ \frac{\tilde{\mathcal V}^2({\bf k})}{\z_{\pm}-\Sigma^{(\pm)}(\varepsilon)- n_\mathrm{i} \tilde{\mathcal V}({\bf k})}
\label{eq:Sigmadscb}
\end{equation}

In contrast to the case of the 
simplest approximation, the equation for $\Sigma$ cannot be analytically solved. In order to simplify
our numerical calculations, we will in the following concentrate on the experimentally relevant regime $\gamma\ll a^3V_0$ (e.g., $\gamma \simeq 0.002~a^3 V_0$ for GaAs). In this case, 
we can neglect the influence of the spin-orbit interaction on the density of states, therefore in Eq.~(\ref{eq:Sigmadscb}) the $2\times 2$-matrix $\tilde{\mathcal V}({\bf k})$  can be replaced by the scalar quantity $\tilde{\mathcal V}_0({\bf k})$. This equation needs to be solved in order to determine $\Sigma^{(+)}(\varepsilon)$ for a given $\varepsilon$. The numerical solution of the self consistent equation can be done by iteration. Starting with $\Sigma^{(+)}(\varepsilon)=-i$ (which verifies the requirement of having a negative imaginary part), successive iterations rapidly converge to a $\Sigma^{(+)}(\varepsilon)$ with negative imaginary part. This self-consistent solution for $\Sigma^{(+)}(\varepsilon)$ yields, through Eqs.~(\ref{eq:dos}) and \eqref{eq:sigmadef}, the density of states presented by dashed lines in Fig.~\ref{fig:rhoeps_and_taueps}(a). This spin-independent $\rho(\varepsilon)$ is the same as the one obtained by Matsubara and Toyozawa \cite{mat-toy} by analytically performing the integral of Eq.~(\ref{eq:Sigmadscb}) and numerically solving the resulting algebraic equation for $\Sigma^{(+)}(\varepsilon)$. 

The sharp cutoff of the LCSCA $\rho(\varepsilon)$ present in Fig.~\ref{fig:rhoeps_and_taueps}(a) is an artifact of the approximation, not vouched by cumulant approach calculations \cite{gas-cyr} and quantum numerical simulations \cite{gibbons_81,chi-hub,proceedings}. 
This is not a serious drawback, since for uncompensated semiconductors the measurable properties, like the spin-relaxation time, only concern the energy of the half-filled band. Moreover, in models that go beyond the Hamiltonian of 
Eqs.~\eqref{eq:Hrest}-\eqref{eq:coupling} 
by incorporating the conduction band, for energies $\varepsilon > V_0$ there will be a hybridization between the tails of the two bands \cite{yonezawa64,Serre,Radjenovic}. The above-mentioned hypothesis of a weak effect of the spin-orbit interaction on $\rho(\varepsilon)$ is validated by quantum numerical calculations using different spin-orbit mechanisms and coupling strengths considerably larger than the realistic ones \cite{proceedings}.

\subsection{Renormalized hopping amplitude matrices}
\label{sec:rha} 

Eq.~\eqref{eq:Sigmadsca} can also be written as
\begin{eqnarray}
\Sigma^{(\pm)}(\varepsilon) &=& 
n_\mathrm{i} G^{(\pm)}(\varepsilon) \int\frac{{\rm d}{\bf k}}{(2\pi)^3} \ {\tilde{\mathcal V}({\bf k})} \ 
\tilde{\mathcal F}^{(\pm)}(\varepsilon,{\bf k}) \nonumber \\
&=&
n_\mathrm{i} G^{(\pm)}(\varepsilon) \int {\rm d}{\bf r} \ {\mathcal V}(-{\bf r}) \ {\mathcal F}^{(\pm)}(\varepsilon,{\bf r}) \, ,
\label{eq:Sigmadscc}
\end{eqnarray}
where the renormalized hopping amplitudes ${\mathcal F}^{(\pm)}(\varepsilon,{\bf r})$ and their Fourier transforms $\tilde{\mathcal F}^{(\pm)}(\varepsilon,{\bf k})$ are defined by
\begin{subequations}
\label{eq:F}
\begin{eqnarray}
\label{eq:F.a}
{\mathcal F}^{(\pm)}(\varepsilon,{\bf r})  &=& \int\frac{{\rm d}{\bf k}}{(2\pi)^3} \ e^{-i{\bf k}\cdot{\bf r}} \ \tilde{\mathcal F}^{(\pm)}(\varepsilon,{\bf k}) \, , \\
\label{eq:F.b}
\tilde{\mathcal F}^{(\pm)}(\varepsilon,{\bf k}) &=& \frac{\tilde{\mathcal V}({\bf k})}{\mathbb{I}_{2} - n_\mathrm{i} \ G^{(\pm)}(\varepsilon) \ \tilde{\mathcal V}({\bf k})} \, ,
\end{eqnarray}
\end{subequations}
respectively. Fig.~\ref{fig:Sigma_loops}(b) shows the diagrammatic expansion leading to the effective hopping ${\mathcal F}^{(\pm)}(\varepsilon,{\bf r})$ (represented by the thick line). The self-energy in the LCSCA displayed in the first line of Fig.~\ref{fig:Sigma_loops}(a) takes the more compact form given in the second line when using the renormalized hopping amplitude.

The para-odd character of $\tilde{\mathcal V}({\bf k})$ is inherited by each term 
$\left[n_\mathrm{i}G^{(\pm)}(\varepsilon) \tilde{\mathcal V}({\bf k})\right]^n$ 
in the expansion leading to Eq.~\eqref{eq:F.b}, and thus also by $\tilde{\mathcal F}^{(\pm)}(\varepsilon,{\bf k})$ and ${\mathcal F}^{(\pm)}(\varepsilon,{\bf r})$. Similarly to Eq.~\eqref{eq:inversion}, we have that ${\mathcal F}^{(\pm)}(\varepsilon,-{\bf r}) \ {\mathcal F}^{(\pm)}(\varepsilon,{\bf r})$ is proportional to $\mathbb{I}_{2}$. 

\begin{figure}[t]
\begin{center}
\includegraphics[width=5cm]{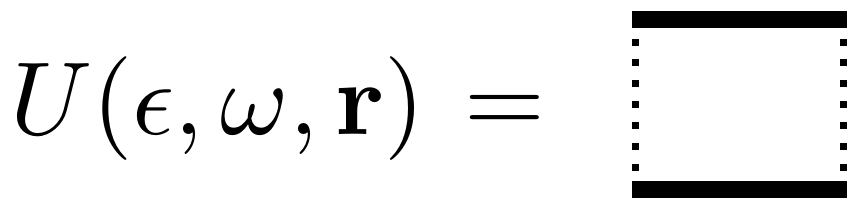}
\caption{
Irreducible component of the intensity propagator within the LCSCA expressed as a renormalized insertion. The thick solid horizontal upper (lower) line stands for the renormalized hopping amplitude matrix ${\mathcal F}^{(+)}(\varepsilon_1,{\bf r})$ (${\mathcal F}^{(-)}(\varepsilon_2,{\bf r})$), the dotted vertical lines indicate identical sites, $\varepsilon_{1,2}=\varepsilon \pm \hbar \omega/2$. 
}
\label{fig:U_effective}
\end{center}
\end{figure}

In analogy with \eqref{eq:F}, we define
\begin{equation}
\tilde{\mathcal F}_0^{(\pm)}(\varepsilon,{\bf k}) = 
\frac{\tilde{\mathcal V}_0({\bf k})}{1-n_\mathrm{i} G^{(\pm)}(\varepsilon) \ \tilde{\mathcal V}_0({\bf k})} \, ,
\label{eq:F0}
\end{equation}
and ${\mathcal F}_0^{(\pm)}(\varepsilon,{\bf r})$ as its inverse Fourier transform. These spin-independent renormalized hopping amplitudes have already been considered in the study of the spatial diffusion within the spinless MT model and its refinements \cite{yonezawa64}. Since $\tilde{\mathcal F}_0^{(\pm)}(\varepsilon,-{\bf k})=\tilde{\mathcal F}_0^{(\pm)}(\varepsilon,{\bf k})$, we also have that $\tilde{\mathcal F}_0^{(\pm)}(\varepsilon,-{\bf r})=\tilde{\mathcal F}_0^{(\pm)}(\varepsilon,{\bf r})$.
The explicit form of ${\mathcal F}_0^{(\pm)}(\varepsilon,{\bf r})$ is given in Appendix B.

\subsection{Irreducible component of the intensity propagator}
\label{sec:icip2} 

Upon application of the recipe for constructing the corresponding irreducible component $U$ of the intensity propagator, the self-energy of Fig.~\ref{fig:Sigma_loops}(a) yields a large proliferation of diagrams. Each $n$-loop term in the expansion gives rise to $n$ contributions to $U$. The renormalized hopping amplitude \eqref{eq:F} is quite useful, as it allows to treat in a systematic way the proliferation of terms. Indeed, the sum of all of them can be compactly expressed through the diagram of Fig.~\ref{fig:U_effective}, which has the same structure as that of Fig.~\ref{fig:irrU}, but where the hopping amplitudes ${\mathcal V}$ have been replaced by the renormalized hopping amplitudes ${\mathcal F}$. The resulting matrix $\tilde{U}(\varepsilon,\omega,{\bf q})$ has a similar expression as that of Eq.~\eqref{eq:A0}, 
\begin{equation}
\tilde{U}(\varepsilon,\omega,{\bf q})  =   n_\mathrm{i}  \int\frac{{\rm d}{\bf k}}{(2\pi)^3} \ \tilde{\mathcal F}^{(+)}(\varepsilon_1,{\bf k}_+)\otimes \tilde{\mathcal F}^{(-)}(\varepsilon_2,{\bf k}_-) \, ,
\label{eq:Fmatrix}
\end{equation}
and allows to determine the charge and spin dynamics within the LCSCA.

\subsection{Spin-relaxation rate}
\label{sec:srr2}

For ${\bf q}=0$, the matrix $\tilde{U}(\varepsilon,\omega,0)$ has the general form of Eq.~(\ref{eq:Atilde}). Using the symmetries of $\tilde{\cal V}({\bf k})$, we can write 
\begin{widetext}
\begin{equation}
\tilde{u}_2(\varepsilon,0)  =   n_\mathrm{i} \int\frac{{\rm d}{\bf k}}{(2\pi)^3} \ \frac{|\tilde{\mathcal C}_x({\bf k})|^2+|\tilde{\mathcal C}_y({\bf k})|^2}
{\left|\bigl(1- n_\mathrm{i} G^{(+)}(\varepsilon)\tilde{\mathcal V}_0({\bf k})\bigr)^2- n_\mathrm{i}  \left[ G^{(+)}(\varepsilon) \right]^2 \left[|\tilde{\mathcal C}_x({\bf k})|^2+|\tilde{\mathcal C}_y({\bf k})|^2+|\tilde{\mathcal C}_z({\bf k})|^2\right]\right|^2}
\label{eq:apm2}
\end{equation}
The $\varepsilon$-dependent spin-relaxation rate follows from inserting this expression into Eq.~(\ref{eq:taus}). Neglecting the term $|\tilde{\mathcal C}_x({\bf k})|^2+|\tilde{\mathcal C}_y({\bf k})|^2+|\tilde{\mathcal C}_z({\bf k})|^2$ in the denominator (due to the condition $\gamma\ll V_0a^3$), and expressing ${\bf k}$ in spherical coordinates, the angular integrals can be performed, yielding 
\begin{equation}
\tilde{u}_2(\varepsilon,0)  = \frac{2^{14} a^6\gamma^2}{105} n_\mathrm{i}
\int\frac{{\rm d}k}{(2\pi)^3} \ 
\frac{k^8}{\left[1+(ka)^2\right]^8} \
\frac{1}
{\left|1- n_\mathrm{i} G^{(+)}(\varepsilon)\tilde{\mathcal V}_0(k) \right|^4} \, .
\label{eq:taulc}
\end{equation}

\end{widetext}

Using the self-consistent $G^{(+)}(\varepsilon)$ obtained from the solution of Eq.~\eqref{eq:Sigmadscb}, and numerically calculating the remaining one-dimensional integral over $k$, results in the spin-relaxation rate $\tau_\mathrm{s}^{-1}(\varepsilon)$ presented in Fig.~\ref{fig:rhoeps_and_taueps}(b) for two values of the impurity density (dashed lines). The $\mathcal{N}_\mathrm{i}$-dependence of the spin-relaxation time $\tau_{\mathrm{s}}(\varepsilon_{\mathrm F})$, for electrons at the Fermi energy in a half-filled impurity band, is shown in Fig.~\ref{fig:agreement} (dashed line). We notice that the LCSCA provides a better description of the numerical results of Ref.~\onlinecite{prl2012} (red dots), as compared with the 
SSCA (dotted line). 
In contrast with our wave-packet \eqref{eq:is}, the numerical determination of the spin-relaxation time used extended initial states (i.e. eigenstates of the spinless problem). Although 
this difference is not expected to strongly
modify the spin-relaxation time in the metallic regime \cite{tam-wei-jal}, it may explain, together with the uncertainty due to numerical finite-size effects, the remaining deviation between theory and numerics in Fig.~\ref{fig:agreement}.
  
\subsection{Spatial diffusion}
\label{sec:sd2}

Neglecting the effect of the spin-orbit coupling for the spatial diffusion,
in accordance with the same approximation ($\gamma\ll a^3 V_0$) used  above for the numerical evaluation of the density of states,
we see that only the matrix element 
${U}^{++,++}(\varepsilon,0,{\bf r})$ contributes in the general expression (\ref{eq:steplength}) of the step-length distribution. We thus obtain
\begin{equation}
p(\varepsilon,{\bf r})=n_{\rm i} \left|G^{(+)}(\varepsilon) {\mathcal F}^{(+)}_0(\varepsilon,{\bf r})\right|^2\ ,
\label{eq:steplengthlosca}
\end{equation} 
where $\tilde{\mathcal F}^{(\pm)}_0(\varepsilon,{\bf k})$ has been defined in \eqref{eq:F0}. We recover, again, a classical random walk,
since $p(\varepsilon,{\bf r})\geq 0$ for all ${\bf r}$.
The step-length distribution results from a superposition of three exponentials with different decay constants, see Eq.~(\ref{eq:f0r}). Such a behavior also differs from that of the step-length distribution predicted by the Boltzmann equation, which contains only a single exponential decay characterized by the scattering mean free path.

The diffusion coefficient emerging from Eqs.~(\ref{eq:diffusionr},\ref{eq:steptime},\ref{eq:steplengthlosca}) is presented as a function of the energy within the impurity band in Fig.~\ref{fig:rhoeps_and_taueps}(c) (dashed lines). These results reproduce those of Ref.~\onlinecite{mat-toy} when using the diffusion approximation for the conductivity $\sigma=e^2\rho(\varepsilon_\mathrm{F}) \ {\cal D}(\varepsilon_\mathrm{F})$. 

\begin{figure}[t]
\begin{center}
\includegraphics[width=8.5cm]{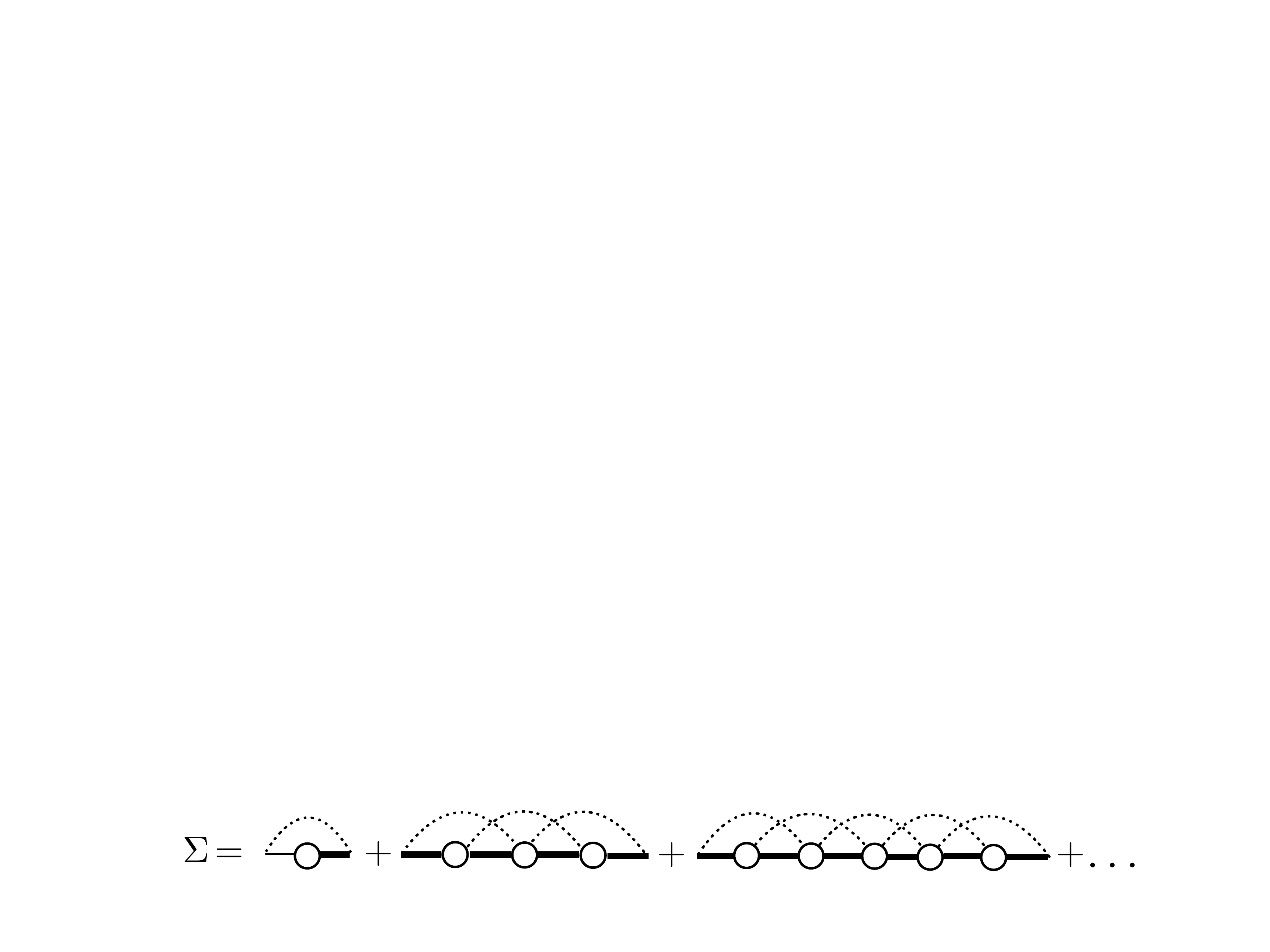}
\caption{Diagrams for the self-energy including repeated scattering.
}
\label{fig:Sigma_RSC-SCA}
\end{center}
\end{figure}

\section{Repeated-scattering-corrected self-consistent approximation}
\label{sec:scasd}

\subsection{Self-energy and density of states}
\label{sec:sedos3}

As discussed in Sec.~\ref{sec:sedos2}, the Matsubara-Toyozawa approximation for the density of states \cite{mat-toy} (equivalent to the splinless version of the LCSCA) gives a relatively good account of the numerically obtained $\rho(\varepsilon)$, up to some deviations in the high-energy part of the impurity band. While it can be argued that these deviations are not significant when considering physically measurable quantities, the search for a more accurate description of the density of states for the spinless version of the model defined by 
Eqs.~\eqref{eq:Hrest}-\eqref{eq:coupling} 
beyond that of Ref.~\onlinecite{mat-toy} is an interesting task \cite{yonezawa64,mat-kan,gas-cyr,chao-oli-majlis,pur-oda_81,gibbons_81,chi-hub}. 

In this section we take the self-consistent approximation to a more complete description by including 
cross 
diagrams 
that describe the repeated scattering from selected 
impurities. Our repeated-scattering-corrected self-consistent approximation (RSCSCA) is mostly relevant for low impurity densities. When compared with the LCSCA, it provides a substantial change of the density of states in the high-energy part of the impurity band. Except for very small impurity densities, it has very little effect on the diffusion coefficient or the spin relaxation rate, thus sustaining the LCSCA results for these two physical quantities.

The simplest way to take into account the repeated scattering from a given set of impurities is to select just a pair. 
Such processes are dominant in the limit of very low impurity densities \cite{Elyutin}.
Therefore, the self-energy of Fig.~\ref{fig:Sigma_loops}(a) can be generalized to that of Fig.~\ref{fig:Sigma_RSC-SCA}, where the hopping amplitudes are the renormalized ones (Fig.~\ref{fig:Sigma_loops}(b) and Eq.~\eqref{eq:F}). 

The expansion for $\Sigma^{(\pm)}(\varepsilon)$ in Fig.~\ref{fig:Sigma_RSC-SCA} starts with the contribution \eqref{eq:Sigmadscc} of the LCSCA, and the following terms represent, for each position ${\bf r}$ of the intermediate impurity, a geometric series with ratio
\begin{equation}
A^{(\pm)}(\varepsilon,{\bf r})= \left[G^{(\pm)}(\varepsilon)\right]^2 \ {\mathcal F}^{(\pm)}(\varepsilon,-{\bf r}) \ {\mathcal F}^{(\pm)}(\varepsilon,{\bf r}) \, ,
\label{eq:defA}
\end{equation}
describing the hopping (with the renormalized hopping amplitude) from one impurity to another one located at distance ${\bf r}$ and back again. As shown in \ref{sec:rha}, ${\mathcal F}^{(\pm)}(\varepsilon,-{\bf r}) \ {\mathcal F}^{(\pm)}(\varepsilon,{\bf r})$ is proportional to the unit matrix, and therefore $A^{(\pm)}(\varepsilon,{\bf r})$ is also a scalar quantity. Summing the geometric series, the self-energy represented in Fig.~\ref{fig:Sigma_RSC-SCA} can be expressed as
\begin{widetext}

\begin{equation}
\Sigma^{(\pm)}(\varepsilon)= n_\mathrm{i} \ G^{(\pm)}(\varepsilon) 
\int{\rm d}{\bf r} 
\left({\mathcal V}(-{\bf r})\ {\mathcal F}^{(\pm)}(\varepsilon, {\bf r})+
\frac{\left[G^{(\pm)}(\varepsilon)\right]^2 
\left[{\mathcal F}^{(\pm)}(\varepsilon,-{\bf r}) \ {\mathcal F}^{(\pm)}(\varepsilon,{\bf r})\right]^2}
{1-A^{(\pm)}(\varepsilon,{\bf r})}\right) \, .
\label{eq:Sigmanew}
\end{equation}  

\end{widetext}

The first term of the integrand, ${\mathcal V}(-{\bf r},\z) \ {\mathcal F}^{(\pm)}(\varepsilon, {\bf r})$, leads to a diagonal contribution upon integration, as can be shown by switching to Fourier space. The second term of the integrand is diagonal, since ${\mathcal F}^{(\pm)}(\varepsilon,-{\bf r}) \ {\mathcal F}^{(\pm)}(\varepsilon,{\bf r})$ and $A^{(\pm)}(\varepsilon,{\bf r})$ are diagonal matrices. 
Therefore the approximation \eqref{eq:Sigmanew} for $\Sigma^{(\pm)}(\varepsilon)$ yields a scalar quantity, in agreement with the general symmetry principles discussed in Sec.~\ref{sec:le}.  

Like in the previous section, we solve the self-consistent equation \eqref{eq:Sigmanew} for $\Sigma^{(+)}(\varepsilon)$ by neglecting the influence of the spin-orbit interaction. Within such an approximation we replace ${\mathcal V}({\bf r})$ by ${\mathcal V}_0({\bf r})$ and ${\mathcal F}^{(+)}(\varepsilon,{\bf r})$ by ${\mathcal F}^{(+)}_0(\varepsilon,{\bf r})$ (the latter defined through Eq.~\eqref{eq:F0} and with an explicit form given by Eq.~(\ref{eq:f0r}) of Appendix \ref{sec:AppendixC}). 
Solving Eq.~\eqref{eq:Sigmanew} by a numerical root solver (or by iteration, which, however, does not converge in all cases), we find that a unique solution for $\Sigma^{(+)}(\varepsilon)$ with negative  imaginary part exists if ${\mathcal N}_{\mathrm i} \geq 0.013$. The resulting density of states, represented by the solid lines in Fig.~\ref{fig:rhoeps_and_taueps}(a), turns out to be correctly normalized, i.e. $\int {\rm d}\varepsilon \rho(\varepsilon)=n_{\mathrm i}$. For ${\mathcal N}_{\mathrm i}<0.013$, we find multiple solutions for $\Sigma^{(+)}(\varepsilon)$ with correspondingly unnormalized densities of states. We therefore 
cannot 
apply the RSCSCA for ${\mathcal N}_{\mathrm i} < 0.013$. This is, however, not a serious drawback, since those values lie below the critical density ${\mathcal N}_{\rm c}=0.017$ mentioned in the introduction, where the non-interacting model adopted in this paper becomes invalid. Furthermore, we note that the low-density limit of the non-interacting model can be addressed by Elyutin's approach \cite{Elyutin}, which is very similar to our RSCSCA, but using the bare hopping amplitude instead of the renormalized one.

Turning back to the density of states shown in Fig.~\ref{fig:rhoeps_and_taueps}(a), we see that differences with respect to the LCSCA $\rho(\varepsilon)$ are noticeable in the high-energy part of the impurity band. The RSCSCA $\rho(\varepsilon)$  results in a smoother maximum and does not exhibit a sharp high-energy cutoff. These features approach the analytical results of the quantum numerical calculations\cite{proceedings,chi-hub}, except for energies $\varepsilon > V_0$. The extension of the impurity band beyond $V_0$ is an unphysical result, as it can be proved that for the spinless version of the model 
\eqref{eq:Hrest}-\eqref{eq:coupling}, all eigenvalues are bounded by $V_0$. However, as it has been argued before, the small imprecisions at the high end of the impurity band are generally unimportant concerning measurable quantities.  

\subsection{Irreducible component of the intensity propagator}
\label{sec:icip3}

When applying to the RSCSCA self-energy of Fig.~\ref{fig:Sigma_RSC-SCA} the recipe for the irreducible component $U$ of the intensity propagator, we have three kinds of contributions. Firstly, those coming from the first term yield the insertion $U$ of Fig.~\ref{fig:U_effective} (diagram (a) in Fig.~\ref{fig:U}). Secondly, there are those obtained by removing one of the Green functions corresponding to the repeatedly visited sites. In particular, the second term in the expansion for $\Sigma^{(\pm)}(\varepsilon)$ yields the diagrams (b)-(d) of Fig.~\ref{fig:U}. The dots after diagram (d) stand for the additional contributions obtained by applying the previous procedure to the subsequent terms of the expansion for $\Sigma^{(\pm)}(\varepsilon)$. We note that the final site is identical to the initial site in diagram (c), but not in (b) and (d). Finally, we have to consider the contributions obtained by removing one of the Green functions appearing inside the renormalized hopping amplitudes (thick lines). Those arising from the second term in the expansion of the self-energy are the diagrams (e)-(h), while the dots after diagram (h) stand for contributions generated by the following terms. In this third kind of diagrams the final site (at the left-hand side of the diagrams)
is not identical to one of the two repeatedly visited sites.  

\begin{widetext}

\begin{figure}
\begin{center}\includegraphics[width=16cm]{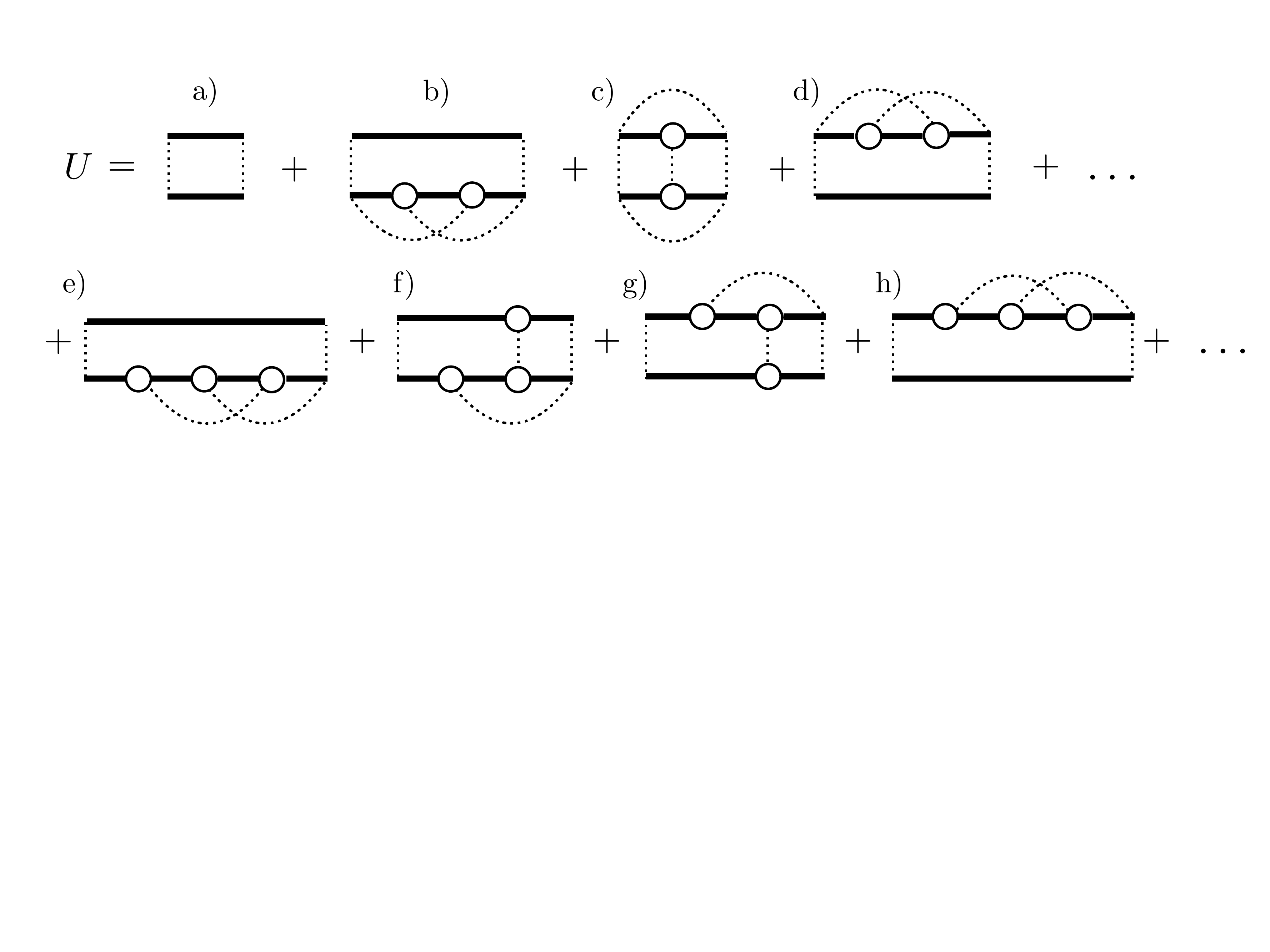}\end{center}
\caption{Diagrams for the irreducible component $U$ of the intensity operator including repeated scattering.\label{fig:U}}
\end{figure}

Diagrams like those of j 
Fig.~\ref{fig:U}(b),(d)
(Fig.~\ref{fig:U}(c)), where the removed Green function does not (does) correspond to the initial site, can be written as

\begin{subequations}
\begin{eqnarray}
{U}_{b-d}^{(n,j)}(\varepsilon,\omega,{\bf r}) 
& = & n_\mathrm{i} 
\left[A^{(+)}\left(\varepsilon_1,{\bf r}\right)\right]^{(j-1)/2}
\left[A^{(-)}\left(\varepsilon_2,{\bf r}\right)\right]^{(n-j)/2}
{\mathcal F}^{(+)}\left(\varepsilon_1,{\bf r}\right) \otimes 
{\mathcal F}^{(-)}\left(\varepsilon_2,{\bf r}\right) \, ,
\label{eq:diagrams_a-b-d}\\
{U}_{c}^{(n,j)}(\varepsilon,\omega,{\bf r}) & = & n_\mathrm{i} \ \delta({\bf r}) \ 
\frac{1}{G^{(+)}\left(\varepsilon_1\right) \ 
G^{(-)}\left(\varepsilon_2 \right)} \
\left[A^{(+)}\left(\varepsilon_1,{\bf r}\right) \right]^{j/2} \
\left[A^{(-)}\left(\varepsilon_2,{\bf r}\right) \right]^{(n+1-j)/2}
\, ,
\label{eq:diagrams_c}
\end{eqnarray}
\end{subequations}
respectively. The odd index 
$n\geq 3$
stands for the number of internal Green functions in the $\Sigma$-diagram (i.e. $n+1$ is order of the diagram), while the odd (even) index $j$ labels the impurity where the cut is applied (with an ordering going from right to left), verifying 
$j \le n$ ($j \le n-1$). 
 For instance, for the diagram of
Fig.~\ref{fig:U}(b) and (d) $(n,j)=(3,1)$ and $(3,3)$, respectively, while for Fig.~\ref{fig:U}(c) $(n,j)=(3,2)$.  
 
The special case of the diagram of Fig.~\ref{fig:U}(a), already treated in Sec.~\ref{sec:icip2}, is accounted for by Eq.~(\ref{eq:diagrams_a-b-d}) when taking $(n,j)=(1,1)$. 

Diagrams  where the removed Green function is within a renormalized hopping amplitude, can be generically written as 

\begin{subequations}
\begin{eqnarray}
{U}_{e-g}^{(n,j)}(\varepsilon,\omega,{\bf r}) & = & 
n_\mathrm{i}^2 G^{(-)}\left(\varepsilon_2 \right) \int {\rm d}{\bf r}'
\left[A^{(+)}\left(\varepsilon_1,{\bf r}'\right)\right]^{(j-1)/2}
\left[A^{(-)}\left(\varepsilon_2,{\bf r}'\right)\right]^{(n-j)/2}
{\mathcal F}^{(+)}\left(\varepsilon_1,{\bf r}\right) \otimes 
{\mathcal F}^{(-)}\left(\varepsilon_2,{\bf r}-{\bf r}'\right)
{\mathcal F}^{(-)}\left(\varepsilon_2,{\bf r}'\right)
\label{eq:diagrams_e_g}\\
{U}_{f-h}^{(n,j)}(\varepsilon,\omega,{\bf r}) & = & 
n_\mathrm{i}^2 G^{(+)}\left(\varepsilon_1\right) \int {\rm d}{\bf r}'
\left[A^{(+)}\left(\varepsilon_1,{\bf r}'\right)\right]^{(j-2)/2}
\left[A^{(-)}\left(\varepsilon_2,{\bf r}'\right)\right]^{(n-j+1)/2}
{\mathcal F}^{(+)}\left(\varepsilon_1,{\bf r}-{\bf r}'\right)
{\mathcal F}^{(+)}\left(\varepsilon_1,{\bf r}'\right) \otimes 
{\mathcal F}^{(-)}\left(\varepsilon_2,{\bf r}\right)
\label{eq:diagrams_f_h}
\end{eqnarray}
\end{subequations}
where the odd index 
$n\geq 3$
again stands for the number of internal Green functions in the $\Sigma$-diagram and $j$ labels the renormalized hopping amplitude where the cut is applied (with an ordering going from right to left). The diagrams of Fig.~\ref{fig:U}(e) and (g) correspond to the case of odd $j$, verifying $j \le n$, i.e. $(n,j)=(3,1)$ and $(3,3)$, respectively. The diagrams of Fig.~\ref{fig:U}(f) and (h) correspond to the case of even $j$, verifying $j \le n+1$, i.e. $(n,j)=(3,2)$ and $(3,4)$, respectively. 

Summing over the allowed values of $n$ and $j$ we obtain the contribution from all diagrams of Fig.~\ref{fig:U}, together with those of higher order. The resulting irreducible component of the intensity propagator in Fourier space is

\begin{eqnarray}
\tilde{U}(\epsilon,\omega,{\bf q}) & = &  n_\mathrm{i}  \int{\rm d}{\bf r} 
\left\{
e^{i{\bf q}\cdot{\bf r}} \ 
\frac{
{\mathcal F}^{(+)}\left(\varepsilon_1,{\bf r}\right) \otimes 
{\mathcal F}^{(-)}\left(\varepsilon_2,{\bf r}\right)
}
{
\left[1-A^{(+)}\left(\varepsilon_1,{\bf r}\right) \right]
\left[1-A^{(-)}\left(\varepsilon_2,{\bf r}\right) \right]
}+\frac{
A^{(+)}\left(\varepsilon_1,{\bf r}\right) \
A^{(-)}\left(\varepsilon_2,{\bf r}\right)
}
{
G^{(+)}\left(\varepsilon_1\right) \ 
G^{(-)}\left(\varepsilon_2 \right)
\left[1-A^{(+)}\left(\varepsilon_1,{\bf r}\right) \right]
\left[1-A^{(-)}\left(\varepsilon_2,{\bf r}\right) \right]
} 
\right\}
+\nonumber\\
& + & 
n_\mathrm{i}^2\int\int {\rm d}{\bf r} \ {\rm d}{\bf r}' e^{i{\bf q}\cdot{\bf r}}  
\left\{ 
\frac{1}
{
\left[1-A^{(+)}\left(\varepsilon_1,{\bf r}'\right) \right]
\left[1-A^{(-)}\left(\varepsilon_2,{\bf r}'\right) \right]
} 
- 1 \right\} \times
\nonumber\\
& &  \Bigl[G^{(-)}\left(\varepsilon_2 \right) 
{\mathcal F}^{(+)}\left(\varepsilon_1,{\bf r}\right) \otimes 
{\mathcal F}^{(-)}\left(\varepsilon_2,{\bf r}-{\bf r}'\right)
{\mathcal F}^{(-)}\left(\varepsilon_2,{\bf r}'\right)
 + G^{(+)}\left(\varepsilon_1\right)  
{\mathcal F}^{(+)}\left(\varepsilon_1,{\bf r}-{\bf r}'\right)
{\mathcal F}^{(+)}\left(\varepsilon_1,{\bf r}'\right) \otimes 
{\mathcal F}^{(-)}\left(\varepsilon_2,{\bf r}\right) 
\Bigr]
\label{eq:Ar}
\end{eqnarray}

\end{widetext}

We notice that $\tilde{U}(\epsilon,\omega,0)$ takes the general form given by Eq.~(\ref{eq:Atilde}).

\subsection{Spin relaxation rate}
\label{subsec:srrrscsca}

According to Eq.~\eqref{eq:taus}, the spin relaxation rate is determined by ${\tilde u}_2(\varepsilon,0)=\tilde{U}^{++,--}(\varepsilon,0,0)$. The 
second contribution to the integrand in the term proportional to $n_\mathrm{i}$ in  Eq.~\eqref{eq:Ar} is a scalar, and therefore does not contribute to ${\tilde u}_2(\varepsilon,0)$. For $\gamma\ll V_0a^3$ the 
first contribution can be expanded up to lowest non-vanishing order in $\gamma$ yielding
\begin{equation}
\frac{{\mathcal F}^{1,-1(+)}\left(\varepsilon,{\bf r}\right) \ 
{\mathcal F}^{1,-1(-)}\left(\varepsilon,{\bf r}\right)
}
{
\left[1-A^{(+)}\left(\varepsilon,{\bf r}\right) \right]
\left[1-A^{(-)}\left(\varepsilon,{\bf r}\right) \right] 
}
\simeq
\left|\frac{{\mathcal F}_{\text{D}}^{(+)}(\varepsilon,{\bf r})}{1-\left[G^{(+)}(\varepsilon)\right]^2 \ \left[{\mathcal F}_{0}^{(+)}(\varepsilon,{\bf r})
\right]^2
}\right|^2 \, ,
\end{equation}
where ${\mathcal F}_{0}^{(\pm)}(\varepsilon,{\bf r})$ has been introduced through Eq.~\eqref{eq:F0} and ${\mathcal F}_{\text{D}}^{(\pm)}(\varepsilon,{\bf r})$ is defined as the inverse Fourier transform of 

\begin{equation}
\tilde{\mathcal F}_{\text{D}}^{(\pm)}(\varepsilon,{\bf k}) =  
\frac{
\pm i\tilde{C}_x({\bf k})+\tilde{C}_y({\bf k})
}{
\left[1- n_\mathrm{i} \ G^{(\pm)}(\varepsilon) \ 
\tilde{\mathcal V}_0({\bf k})\right]^2} \, .
\label{eq:FD}
\end{equation}
${\mathcal F}_{\text{D}}^{(\pm)}(\varepsilon,{\bf r})$  can be calculated analytically, as described in Appendix \ref{sec:AppendixC}.
Keeping also the lowest non-vanishing order in $\gamma$ for the contribution to the term proportional to $n_\mathrm{i}^2$ in  Eq.~\eqref{eq:Ar} we have

\begin{widetext}

\begin{eqnarray}
{\tilde u}_2(\varepsilon,0) & = &  n_\mathrm{i} \int{\rm d}{\bf r} 
\left|\frac{{\mathcal F}_{\text{D}}^{(+)}(\varepsilon,{\bf r})}{1-\left[G^{(+)}(\varepsilon)\right]^2 \ \left[{\mathcal F}_{0}^{(+)}(\varepsilon,{\bf r})
\right]^2
}\right|^2
+ n_\mathrm{i}^2 \int{\rm d}{\bf r}\left\{
\left|\frac{1}
{1-\left[G^{(+)}(\varepsilon)\right]^2 \ \left[{\mathcal F}_{0}^{(+)}(\varepsilon,{\bf r})
\right]^2
}\right|^2
-1\right\} \times
\nonumber\\
& & \times\int\frac{{\rm d}{\bf k}}{(2\pi)^3} \ 
2{\rm Re}\left\{e^{-i{\bf k}\cdot{\bf r}} \ 
G^{(-)}(\varepsilon) \
\tilde{\mathcal F}^{(+)}_{\text{D}}(\varepsilon,{\bf k}) \left[
\tilde{\mathcal F}_{0}^{(-)}(\varepsilon,{\bf k}) \
{\mathcal F}^{(-)}_{\text{D}}(\varepsilon,{\bf r}) + 
\tilde{\mathcal F}_{\text{D}}^{(-)}(\varepsilon,{\bf k}) \ 
{\mathcal F}_{0}^{(-)}(\varepsilon,{\bf r})\right]\right\} \, .
\label{eq:Apmspin}
\end{eqnarray}

Using the explicit forms of ${\mathcal F}_{0}^{(\pm)}(\varepsilon,{\bf r})$ and ${\mathcal F}_{\text{D}}^{(\pm)}(\varepsilon,{\bf r})$, it is possible to analytically perform in Eq.~(\ref{eq:Apmspin}) the ${\bf k}$-integral as well as the angular part of the ${\bf r}$-integral (although the resulting expressions, which we obtained using a symbolic calculus computer program, are too long to be displayed). Numerically performing the remaining integral over $r$, the resulting spin-relaxation rate is shown, as a function of the energy $\varepsilon$, by the solid line in  Fig.~\ref{fig:rhoeps_and_taueps}(b) for two chosen impurity densities. The spin-relaxation time for electrons at the Fermi energy $\tau_{\mathrm{s}}(\varepsilon_{\mathrm F})$ is presented in Fig.~\ref{fig:agreement} (solid line). Except for low impurity densities, where we expect the RSCSCA to be most relevant, the differences for the spin-relaxation results between the LCSCA and the RSCSCA are quite small. Such finding is important in order to sustain the validity of the former. 
The fact that improving the loop-corrected self-consistent approximation  (valid in the limit of high densities) by the consideration of terms 
that give the leading contributions at low densities
does not significantly alter the resulting spin relaxation rate,
confirms that the high density-limit described by the LCSCA is indeed applicable in the impurity density interval of interest.

\subsection{Spatial diffusion coefficient}
\label{subsec:sdrscsca}

Neglecting the effect of the spin-orbit coupling for the spatial diffusion leads to 

\begin{eqnarray}
p(\varepsilon,{\bf r})& = & n_\mathrm{i}\left|G^{(+)}(\varepsilon)\right|^2\left\{
\left|\frac{{\mathcal F}_{0}^{(+)}(\varepsilon,{\bf r})}{1-\left[G^{(+)}(\varepsilon)\right]^2 \ \left[{\mathcal F}_{0}^{(+)}(\varepsilon,{\bf r})
\right]^2
}\right|^2
+\delta({\bf r})\int {\rm d}{\bf r}'  \left|\frac{G^{(+)}(\varepsilon) \ \left[{\mathcal F}_{0}^{(+)}(\varepsilon,{\bf r}')\right]^2}{1-\left[G^{(+)}(\varepsilon)\right]^2 \ \left[{\mathcal F}_{0}^{(+)}(\varepsilon,{\bf r}')
\right]^2}\right|^2
\right.+\nonumber\\
& & 
\!\!\!\!\!\!\! \!\!\!\!\!\!\! \left.
+n_\mathrm{i}\int {\rm d}{\bf r}' 
\left[
\left|\frac{1}
{1-\left[G^{(+)}(\varepsilon)\right]^2 \ \left[{\mathcal F}_{0}^{(+)}(\varepsilon,{\bf r})
\right]^2
}\right|^2
-1\right] \
2{\rm Re}\left[G^{(-)}(\varepsilon) \ 
{\mathcal F}_{0}^{(+)}(\varepsilon,{\bf r}) \ 
{\mathcal F}_{0}^{(-)}(\varepsilon,{\bf r}') \ 
{\mathcal F}_{0}^{(-)}(\varepsilon,{\bf r}-{\bf r}')\right]
\right\} \, ,
\label{eq:DRSCSCA}
\end{eqnarray}

\end{widetext}
from which the diffusion constant follows through Eqs.~(\ref{eq:diffusionr},\ref{eq:steptime}). The result is
plotted in Fig.~\ref{fig:rhoeps_and_taueps}(c) as a function of $\varepsilon$ (solid lines). Like in the case of the spin relaxation rate, the diffusion constant of the RSCSCA is very close to that of the LCSCA. 
In contrast to the previous approximations, $p(\varepsilon,{\bf r})$ is not manifestly positive, and therefore it does not correspond to a step-length probability distribution. Indeed, we find (in the relevant regime of densities shown in Fig.~\ref{fig:agreement}) that the function $p(\varepsilon,{\bf r})$ takes negative values for some distances ${\bf r}$, especially for energies close to the band edges. Hence, the spatial dynamics predicted by the RSCSCA cannot be interpreted, in general, as a classical random walk.

\section{Conclusion}
\label{sec:conclusion}

The self-consistent approximation for non-interacting electrons in disordered systems \cite{vol-woe,rev-vol-woe,kkw90} has been generalized as to include the effect of spin-orbit interaction in a random  
network model describing the impurity band of doped semiconductors with zinc-blende crystal structure. The model is relevant for impurity densities larger than the critical one for the metal-insulator transition, where electron conduction takes place in the impurity band. The case of the Dresselhaus spin-orbit coupling has been considered, since it has been proven to be the source of the dominant spin-relaxation mechanism in a wide class of materials \cite{tam-wei-jal,prl2012}, accounting for the experimentally measured values \cite{kik-aws,dzh,oes-roe-hau-hae,sch-hei-roh,roe-ber-mue,spr-etal}. But the approach presented in this article can be generalized to other spin-orbit coupling mechanisms, as well as to different crystal structures. 

The inclusion of spin-orbit interaction in the Matsubara-Toyozawa random 
network model leads to the introduction of a hopping amplitude matrix, which for the case of zinc-blende symmetry has very special transformation properties. In the footsteps of Vollhardt and W\"olfle \cite{vol-woe,rev-vol-woe} we have provided the recipe for building diagrams of the irreducible component of the intensity propagator from those of the irreducible self-energy. Such a procedure leads to a Ward identity ensuring the required conservation laws at each level of approximation in the self-consistent scheme. 

The link between one- and two-particle quantities allowed us to obtain the density of states in the impurity band, as well as the energy-dependent diffusion constant and spin-relaxation time. Analytical and semi-analytical expressions of these quantities could be reached for the case of the simpler self-consistent schemes. The spin-coherence and spin-relaxation times coincide for the considered case of zinc-blende symmetry. The spin-orbit corrections to the diffusion constant are very small in the regime of weak spin-orbit coupling. 
Similarly, the spin-orbit corrections to the density of states are extremely weak, and can generically be ignored \cite{proceedings}.

Describing the energy-dependent diffusion coefficient and spin-relaxation time allows to address not only the case of uncompensated semiconductors (with a half-filled impurity band), but also that of an arbitrary degree of compensation. While in this work we have concentrated our attention to the case of a weak optical excitation, where the carrier density is fixed by the doping, experiments with an elevated optical excitation condition \cite{schneider10} can be analyzed using our results for the energy-dependent spin-relaxation. 

The simplest self-consistent scheme evaluates the local Green function in terms of processes where the electron hops to another impurity, and then back again. It  yields a 'semi-circle law' for the energy-dependence of the density of states, the diffusion coefficient, and the spin-relaxation rate. The value of the latter at the Fermi energy qualitatively reproduces the phenomenological results of Ref.~\onlinecite{prl2012}.

Taking into account round trips of the electron including more than one impurity 
while neglecting multiple visits to a given impurity (i.e. the so-called cross diagrams in the perturbation expansion) leads in the spinless case to the well-studied  
Matsubara-Toyozawa approach applied to their random model 
(that we referred to as 'loop-corrected self-consistent approximation'). 
This approximation provides the leading contribution in the limit of high impurity density.
Important corrections in the density of states and the diffusion coefficient appear, with respect to the 
simplest
self-consistent approximation, as well as a reduction of the spin-relaxation time by a factor of $2$, improving the agreement with existing numerical simulations \cite{prl2012}. The agreement with the experimental results is qualitative, in view of the uncertainty with which the value of the Dresselhaus spin-orbit coupling constant is known \cite{eng-ras-hal,fab-etal,car-chr-fas,cha-van-kot,kri-hal,mar-ste-tit,knap-etal,mei-etal,jusserand}.

The treatment of cross diagrams in a simplified way  (i.e. taking into account only pairs of impurities), 
referred to as 'repeated-scattering corrected self-consistent approximation', improves the results of the previous approximation by providing small  corrections  to the spin relaxation rate and the diffusion constant. These diagrams  give the leading contribution in the limit of low impurity density \cite{Elyutin},
where the repeated scattering off a given impurity becomes more likely, and are therefore expected to yield the most important corrections with respect to the above  high-density limit. While in a diagrammatic expansion it is always difficult to prove the convergence, the fact that  these corrections are small hence confirms the validity of the loop-corrected approximation in the regime of impurity densities ($n_{\rm c}<n_{\rm i}<n_{\rm h}$) that we are interested in.  

As a byproduct of our study aiming the spin relaxation in doped semiconductors, the self-consistent approximation provides a useful and systematic path for understanding the spinless case of random 
network models. Various theoretical schemes, developed in different contexts  \cite{mat-toy,Elyutin,yonezawa64,mat-kan,gas-cyr,pur-oda_81,gibbons_81,chi-hub,Serre, proceedings} have been discussed in a unified way under the light of a self-consistent scheme that links one- and two-particle physical quantities and guarantees probability conservation at each level of approximation. The interest in studying these models is not only restricted to the description of the impurity band in doped semiconductors, but it applies to a variety of physical contexts, e.g. in order to characterize diffusion of excitations mediated by resonant dipole-dipole interactions  
in ultracold Rydberg gases \cite{Rydberg_ex,Rydberg_th}.

Our work opens the perspective of treating other crystal structures, like wurtzite, where experiments on spin coherence in GaN \cite{beschoten01} have not received so far a consistent theoretical description. In addition, a more elaborate treatment of the cross diagrams would allow to obtain reliable results for densities approaching the critical one from the metallic side of the transition. Some of the theoretical schemes used to include electron-electron interactions in the random lattice model \cite{Serre,slevin14}
or the recently developed diagrammatic methods for treating interactions between quantum particles propagating in random potentials \cite{geiger12,geiger13},
could be adapted within the self-consistent framework in order to approach the transition within a reliable model.

\acknowledgments

R.A.J. thanks P. Tamborenea, D. Vollhardt, and D. Weinmann for fruitful discussions; he also acknowledges financial support from the French National Research Agency ANR (Project Labex NIE) and the French-Argentinian collaborative project PICS 06687. The research leading to these results has received funding from the People Programme (Marie Curie Actions) of the European Union's Seventh Framework Programme (FP7/2007- 2013) under REA Grant agreement no. [609305]. 

\appendix

\section{Charge and spin dynamics from the intensity propagator}
\label{sec:AppendixA}

In this appendix we proof Eq.~(\ref{eq:defPrtssp}) of the main text, closely following  chapter 4.1 of Ref.~\onlinecite{akk-mont}. We first determine the normalization constant $A$ in Eq.~(\ref{eq:is}):
\begin{eqnarray}
\label{eq:normalization}
1 &=& |A|^2 \overline{ \sum_{\nu} \ \left| \langle \chi_{\nu} \ | \ m \sigma \rangle \right|^2 \ \exp{\left[-\frac{\left(\varepsilon_{\nu}-\varepsilon\right)^2}{2\sigma_{\varepsilon}^2}\right]}} \nonumber \\
&=& \frac{|A|^2}{n_\mathrm{i}} \int_{-\infty}^{+\infty} {\rm d}\tilde{\varepsilon} \
\rho(\tilde{\varepsilon}) \ \exp{\left[-\frac{\left(\tilde{\varepsilon}-\varepsilon\right)^2}{2\sigma_{\varepsilon}^2}\right]} \nonumber \\
&=& \frac{|A|^2}{n_\mathrm{i}} \ \sqrt{2\pi}\sigma_{\varepsilon} \ \rho(\varepsilon)
\, ,
\end{eqnarray}
where the density of states $\rho(\varepsilon)$ is assumed to be slowly-varying in the scale of $\sigma_{\varepsilon}$, and the normalization is assumed to hold only on average. Then, we note that:
\begin{equation}
\langle m'\sigma'|{\cal U}_t|\psi_{\epsilon,m,\sigma}\rangle=i A\int_{-\infty}^{+\infty} \frac{{\rm d}\tilde{\varepsilon}}{2\pi} \
g_{m'm}^{\sigma'\sigma(+)}(\tilde{\varepsilon}) \
e^{-\frac{(\tilde{\varepsilon}-\varepsilon)^2}{4\sigma_\varepsilon^2}}e^{-i\tilde{\varepsilon}t/\hbar} \, ,
\end{equation}
what can be deduced by using the definitions of the Green function, Eq.~(\ref{eq:smallg}), and of the 
initial state, Eq.~(\ref{eq:is}). Inserting this expression into Eq.~(\ref{eq:Pphysical}), we get:
\begin{eqnarray}
P^{\sigma'\sigma}(\varepsilon,t,{\bf r})  & = &  \hbar |A|^2 \int_{-\infty}^{+\infty}  \frac{{\rm d} \tilde{\varepsilon}}{2\pi}
\int_{-\infty}^{+\infty} \frac{{\rm d}\omega}{2\pi} \
\Phi^{\sigma'\sigma',\sigma\sigma}(\tilde{\varepsilon},\omega,{\bf r}) \
\nonumber \\
\ \ & \times &
e^{-i\omega t} \ e^{-\frac{(\tilde{\varepsilon}_1-\varepsilon)^2}{4\sigma_\varepsilon^2}} e^{-\frac{(\tilde{\varepsilon}_2-\varepsilon)^2}{4\sigma_\varepsilon^2}} \, ,
\label{eq:C3}
\end{eqnarray}
with $\tilde{\varepsilon}_{1,2}=\tilde{\varepsilon}\pm\hbar\omega/2$. The integration over $\tilde{\varepsilon}$ can now be performed by assuming that the impurity-averaged intensity propagator $\Phi^{\sigma'\sigma',\sigma\sigma}(\tilde{\varepsilon},\omega,{\bf r})$ does not strongly depend on
$\tilde{\varepsilon}$ (on the scale given by $\sigma_\varepsilon$), and can therefore be taken outside the integral:
\begin{equation}
|A|^2 \int_{-\infty}^{+\infty} {\rm d} \tilde{\varepsilon}~e^{-\frac{(\tilde{\varepsilon}_1-\varepsilon)^2}{4\sigma_\varepsilon^2}} e^{-\frac{(\tilde{\varepsilon}_2-\varepsilon)^2}{4\sigma_\varepsilon^2}}\simeq |A|^2 \sqrt{2\pi}\sigma_\varepsilon =\frac{n_{\rm i}}{\rho(\varepsilon)}
\end{equation}
for $\omega\ll \sigma_{\varepsilon}$. This concludes the derivation of Eq.~(\ref{eq:defPrtssp}).

\section{Technical aspects of the self-consistent approximation for the spin-dependent locator expansion}
\label{tascale}

The self-consistent diagrammatic theory developed by Vollhardt and W\"olfle
\cite{vol-woe,rev-vol-woe} for the Edwards model \cite{edwards} describes, for the spinless case, the scaling behavior in the vicinity of the metal-insulator transition, which is driven by disorder in the continuum. The extension to the lattice case has been  carried out, within a locator expansion, by Kroha, Kopp, and W\"olfle \cite{kkw90}. The self-consistent treatment of the particle-particle (cooperon) contributions, together with the multiple-occupancy corrections, yielded for the Anderson model a phase diagram as a function of energy and disorder that is in quantitative agreement with the results of numerical diagonalization of finite-size systems with different disorder distributions (i.e. box-like, Gaussian and Lorentzian). 

The Matsubara-Toyozawa model (the spinless version of the system defined in 
Eq.~\eqref{eq:Hrest}-\eqref{eq:coupling}) 
-- which we also refer to as "random network model"  --
can be thought of as an Anderson model with binary disorder (since, at each point in space, an impurity may be present or not).
The locator expansion has proven in this situation to yield results equivalent to those obtained by using the Bloch states as the basis for the perturbation, provided all irreducible diagrams are included and the multiple occupancy corrections are made self-consistently \cite{Leath}. This equivalence is however difficult to exploit in order to compute specific physical quantities. Our approach in this article has been to restrict ourselves to a selected class of diagrams present in the locator expansion, verifying that the conservation laws are respected at each level of approximation. Particle conservation imposes restrictions to the changes in space and time of the electron density, and taking these restrictions into account is one of the bases of the self-consistent approach.  

The introduction of spin-orbit interaction in the continuum model allowed to address the spin relaxation in disordered metals and heavily doped semiconductors by standard diagrammatic techniques built in momentum space \cite{altshuler82,sachdev87,raimondi90}. The need to describe the impurity band forces us to take a different starting point and adopt a discrete model of randomly distributed impurities, where a locator expansion can be implemented.

In this appendix we show how some of the basic concepts of the self-consistent approach translate into the spin-dependent model 
\eqref{eq:Hrest}-\eqref{eq:coupling}. 
In particular, we give examples of reducible and irreducible diagrams, we provide the recipe for constructing irreducible $U$-diagrams from irreducible $\Sigma$-diagrams, and we prove that such a prescription leads to the Ward identity \eqref{eq:ward0}.

\subsection{Reducible and irreducible diagrams}
\label{apprid}

\begin{figure}
\begin{center}
\includegraphics[width=7cm]{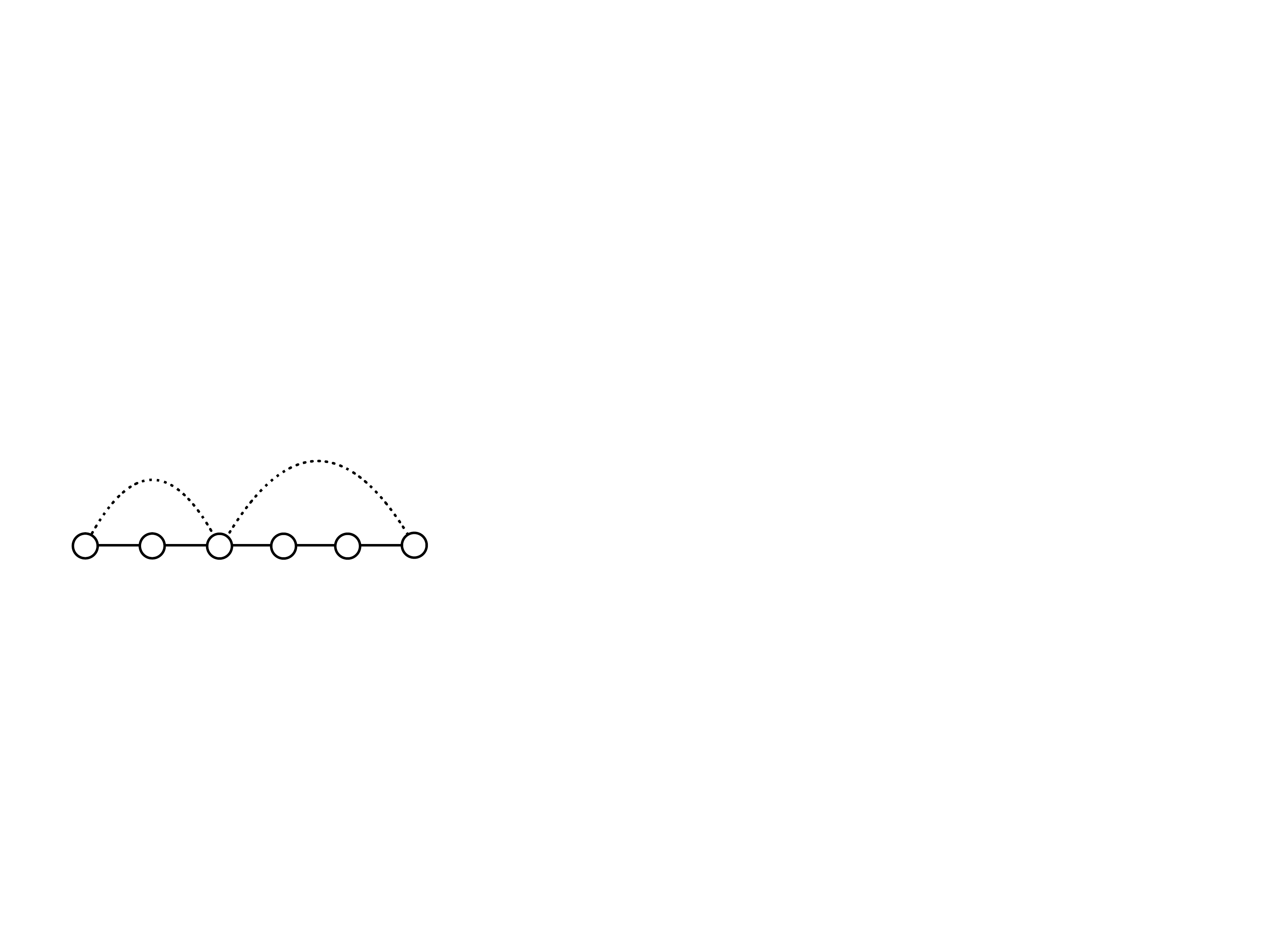}
\caption{Example of a reducible diagram contributing to the average local Green function. The solid horizontal lines represent the hopping matrix ${\cal V}$, the circles stand for the Green functions, and dotted lines indicate identical sites.\label{fig:fig_app_reducible}}
\end{center}
\end{figure}

Diagrams contributing to $G^{(\pm)}(\varepsilon)$ that when 'cut' at an intermediate Green function $G_{00}^{(\pm)}(\varepsilon)$ (or $G^{(\pm)}(\varepsilon)$ if we are working in a self-consistent approach) result in two {\it disconnected} diagrams are defined as reducible. For instance, the fifth-order diagram with three identical sites in Fig.~\ref{fig:fig_app_reducible} is reducible, since cutting it at the third dot, yields two lower-order disconnected diagrams (both of them irreducible, and the first one being identical to diagram Fig.~\ref{fig:fig1_GF}(a)). A similar definition applies to the diagrams corresponding to $\Sigma^{(\pm)}(\varepsilon)$.

\begin{figure}
\begin{center}\includegraphics[width=8cm]{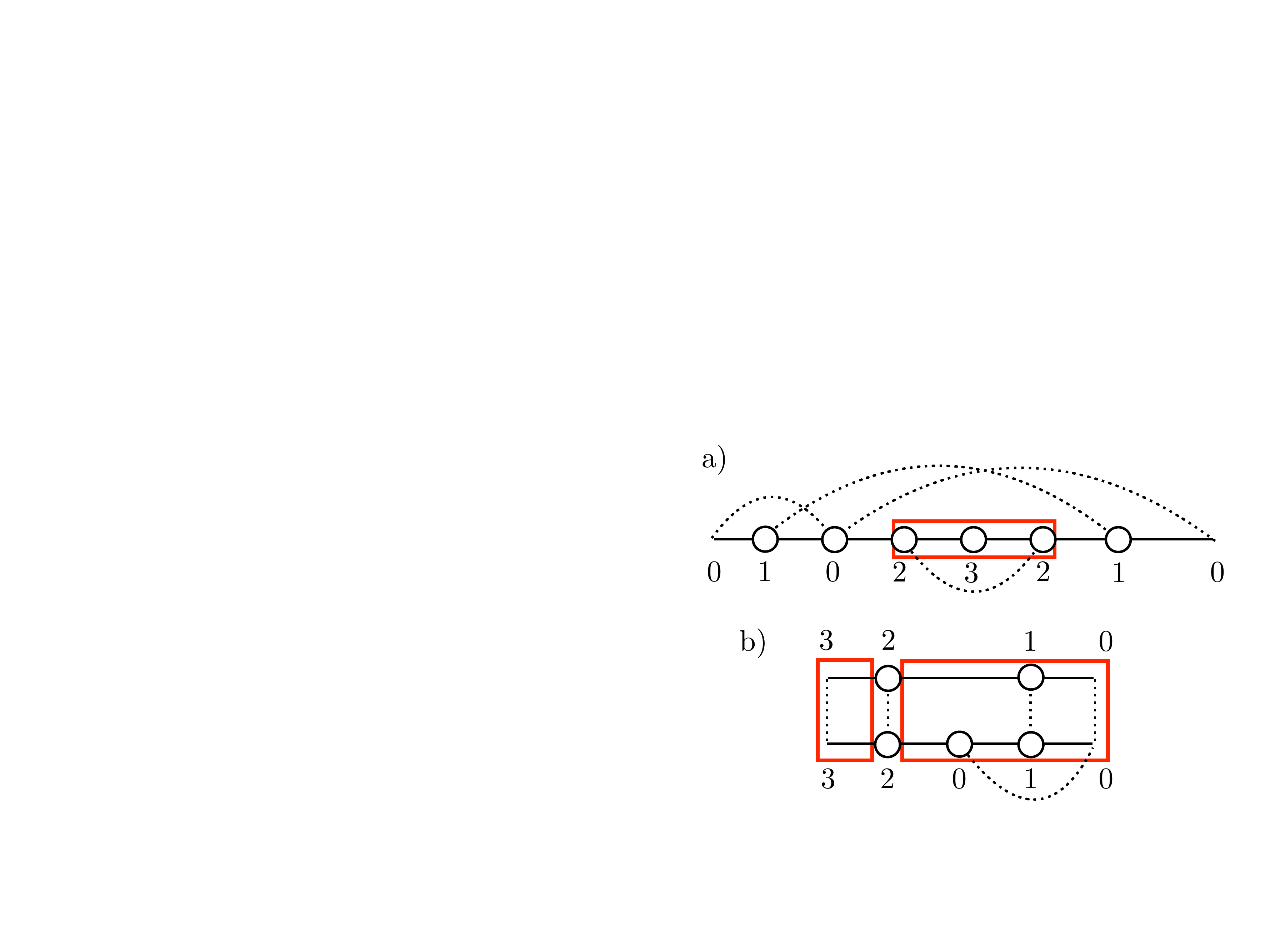}\end{center}
\caption{(a) Example of a two-point reducible diagram. Since the central part (marked by the red rectangle) can be seen as part of  the average local Green function at impurity $2$, (see Fig.~\ref{fig:fig1_GF}(a)),
this diagram is effectively contained in the diagram displayed in Fig.~\ref{fig:cross-a-b}(a) below, and must therefore {\em not} be counted again when evaluating $\Sigma$ in the self-consistent approach. 
(b) Applying the recipe for constructing $U$-diagrams from the $\Sigma$-diagram of (a) -- with cut at the impurity $3$ -- yields a {\em reducible} intensity diagram, which can be expressed as the product of Green functions and two irreducible insertions (marked by the two red rectangles).
\label{fig:cross-c-d}}
\end{figure}

\begin{figure}
\begin{center}\includegraphics[width=8cm]{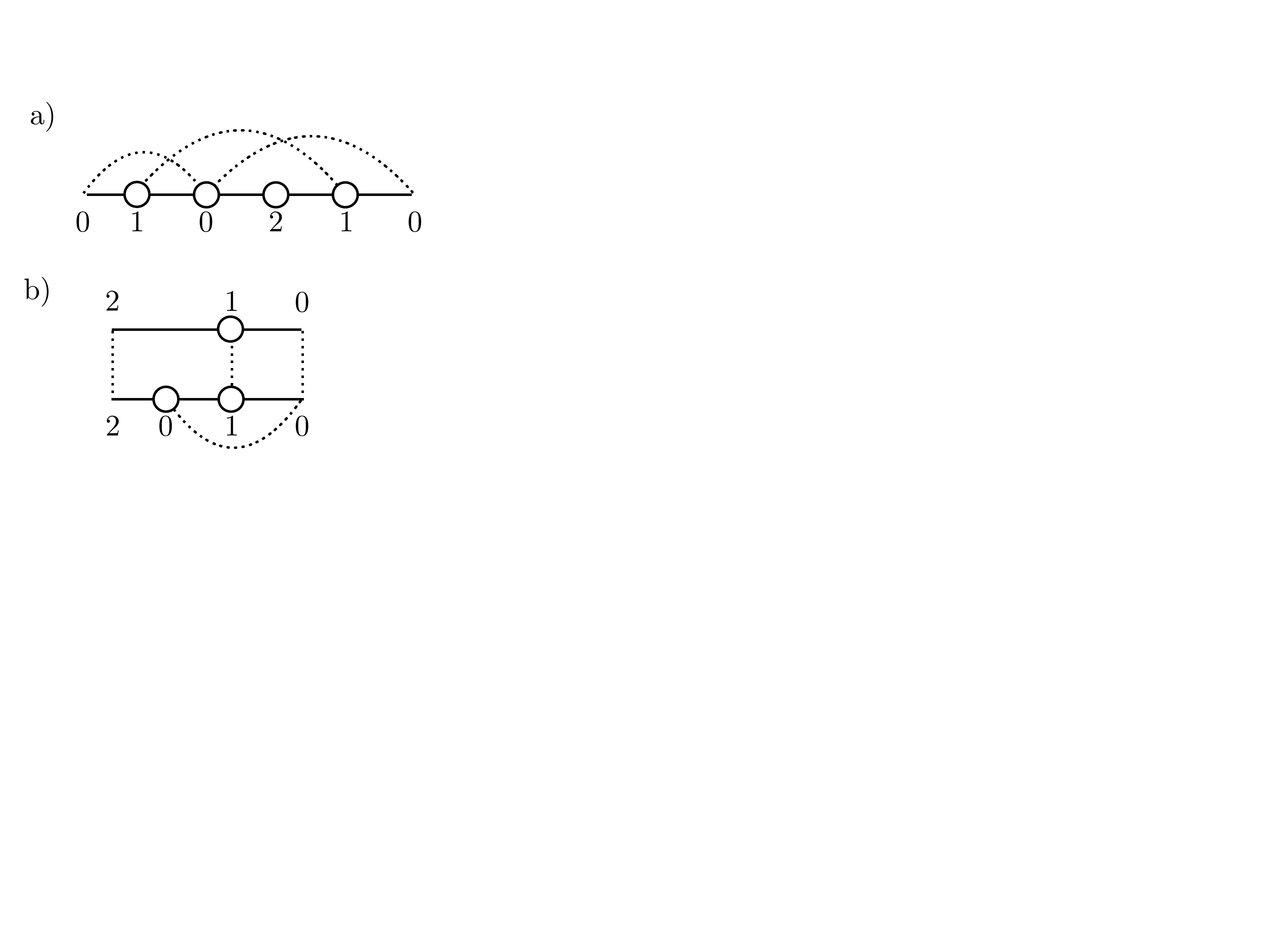}\end{center}
\caption{(a) Example of a diagram $\Sigma_{\bf i}$ for the local self-energy, characterized by the labels ${\bf i}=\{1,2,0,1\}$ of the intermediate impurities (from right to left), see Eq.~(\ref{eq:sigmaward}). b) Corresponding diagram 
$U_{{\bf i},2}$
contributing to the intensity operator, obtained according to the construction recipe described in the text by cutting the $\Sigma$-diagram (a) at impurity $2$.
\label{fig:cross-a-b}}
\end{figure}

The discussion of the irreducibility requires the indexation of the impurities within the diagrams. Considering the locator expansion \eqref{eq:serieszH} in its self-consistent version, we notice that each diagram contributing to the self-energy $\Sigma^{(\pm)}(\varepsilon)$ can be characterized by a set of indices ${\bf i}=\{i_1,\dots,i_n\}$, where $n$ is the number of local Green functions occurring in the diagram (equivalently, $n+1$ is the order of the diagram), and $i_{j}$ labels the impurity corresponding to the $j$-th Green function (ordered from right to left). Identical impurities (which are connected by a dotted line in the corresponding diagram) carry the same label. We note $l$ the number of impurities different from the initial and final one ($l \le n$). For the initial impurity (which is identical to the last impurity), we choose the label $0$, and $1,2,\dots,l$ for the remaining impurities (i.e. we always have $i_1=1$).  As examples of the chosen notation, the diagram shown in Fig.~\ref{fig:cross-c-d}(a) corresponds to
${\bf i}=\{i_1,i_2,i_3,i_4,i_5,i_6\}=\{1,2,3,2,0,1\}$, with $n=6$ and $l=3$, while that of Fig.~\ref{fig:cross-a-b}(a) corresponds to ${\bf i}=\{i_1,i_2,i_3,i_4\}=\{1,2,0,1\}$, with $n=4$ and $l=2$.

The {\it two-point irreducibility} presented in Sec.~\ref{subsec:icotpgf} in order to avoid double counting of diagrams can be addressed with the suggested notation. Fig.~\ref{fig:cross-c-d}(a) presents an example of a diagram that does not contribute to the self-consistent $\Sigma$, since cutting it in the two places where impurity $2$ appears leads to two lower-order unconnected diagrams (the central one, marked by the red rectangle, corresponding to the self-energy contribution of Fig.~\ref{fig:fig1_GF}(b) after removing the dots at impurity $2$). Thus, such a reducible diagram is accounted for in the expansion of the Green function, provided that in the retained approximation for the self-energy the two lower-order diagrams were included. The diagram of Fig.~\ref{fig:cross-a-b}(a) is irreducible, since we do not obtain disconnected diagrams either by cutting it at any of its intermediate impurities, neither by cutting it at the two places where the site $1$ (repeated in the sequence of visited impurities) appears. 

\begin{figure}
\begin{center}
\includegraphics[width=5cm]{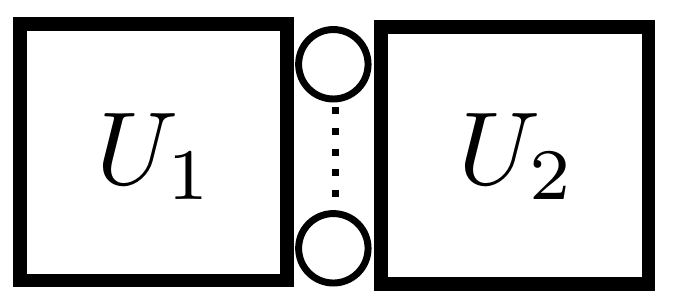}
\caption{Example of a reducible component of the intensity propagator expressed as the product of Green functions representing the same impurity
and two irreducible insertions $U_1$ and $U_2$.
}
\label{fig:factorize}
\end{center}
\end{figure}

The irreducibility of a diagram contributing to the intensity propagator is defined by the property of not being able to be factorized into two components $U_1$ and $U_2$ with an intermediate pair of Green functions representing the same impurity. Fig.~\ref{fig:factorize} shows a counterexample with an insertion of the intensity propagator which is not irreducible.

\subsection{Recipe for constructing diagrams contributing to the  irreducible component of the intensity propagator}
\label{apprbicip}

In the self-consistent approach of Vollhardt and W\"olfle \cite{vol-woe,rev-vol-woe} applied to the Edwards model, the irreducible component of the intensity propagator is generated in a systematic way by taking functional derivatives of the self-energy with respect to each of the Green functions appearing in the corresponding diagram. In a diagrammatic
formulation this recipe amounts to a 'cut and fold' procedure, which we implement in the locator expansion by defining the following steps:

(i) Cut the retarded $\Sigma$-diagram at each of its local Green functions. 

(ii) Take the right-hand part (with respect to this cut) of the self-energy diagram as the upper line of the intensity diagram (at energy $\varepsilon_1$). 

(iii) Fold the rest of the diagram (at the left of the cut) into the lower line of the intensity diagram (at energy $\varepsilon_2$) by taking its complex-conjugate and reversing the spatial coordinate. 

(iii) Add the contributions obtained for each cut.

The retarded self-energy contribution corresponding to the diagram defined by the set ${\bf i}$ writes

\begin{widetext}

\begin{equation}
\Sigma_{{\bf i}}^{(+)}(\varepsilon)=n_\mathrm{i}^l \
\left[G^{(+)}(\varepsilon)\right]^n \
\int{\rm d}{\bf r}_1\dots{\rm d}{\bf r}_l 
{\mathcal V}({\bf r}_0-{\bf r}_{i_{n}}) 
{\mathcal V}({\bf r}_{i_{n}}-{\bf r}_{i_{n-1}})\dots
{\mathcal V}({\bf r}_{i_2}-{\bf r}_{1}) {\mathcal V}({\bf r}_{1}-{\bf r}_0) 
\, .
\label{eq:sigmaward}
\end{equation}

The application of the "cut and fold" procedure to $\Sigma_{{\bf i}}^{(+)}(\varepsilon)$ at each intermediate Green function corresponding to the site $j$ generates the irreducible contribution
\begin{equation}
U_{{\bf i}}(\varepsilon,\omega,{\bf r})=\sum_{j=1}^{n}U_{{\bf i},j}(\varepsilon,\omega,{\bf r}) \, .
\label{eq:usumward}
\end{equation}
Two cases have to be distinguished. If the place where the self-energy was cut corresponds to the initial impurity ($i_j=0$), then
\begin{eqnarray}
U_{{\bf i},j}(\varepsilon,\omega,{\bf r}) & = & n_\mathrm{i}^l \
\delta({\bf r})
\left[G^{(+)}\left(\varepsilon_1\right)\right]^{j-1}
\left[G^{(-)}\left(\varepsilon_2\right)\right]^{n-j}
\int{\rm d}{\bf r}_1\dots{\rm d}{\bf r}_l \nonumber\\
& & {\mathcal V}({\bf r}_0-{\bf r}_{i_{j-1}}) 
\dots {\mathcal V}({\bf r}_{i_2}-{\bf r}_{1})
{\mathcal V}({\bf r}_{1}-{\bf r}_{0})
 \otimes 
 {\mathcal V}^*({\bf r}_{0}-{\bf r}_{i_{j+1}}) \dots 
{\mathcal V}^*({\bf r}_{i_{n-1}}-{\bf r}_{i_{n}}) 
 {\mathcal V}^*({\bf r}_{i_{n}}-{\bf r}_{0}) \, . \label{eq:uward2}
\end{eqnarray}
While if the cut is done at an impurity which is different from the initial one ($i_j\neq 0$), we have
\begin{eqnarray}
U_{{\bf i},j}(\varepsilon,\omega,{\bf r}_{i_j}-{\bf r}_0) & = &
n_\mathrm{i}^l \
\left[G^{(+)}\left(\varepsilon_1\right)\right]^{j-1}
\left[G^{(-)}\left(\varepsilon_2\right)\right]^{n-j}
\int{\rm d}{\bf r}_1\dots{\rm d}{\bf r}_{i_j-1}{\rm d}{\bf r}_{i_j+1}\dots {\rm d}{\bf r}_l \nonumber\\
& & 
{\mathcal V}({\bf r}_{i_{j}}-{\bf r}_{i_{j-1}}) 
\dots {\mathcal V}({\bf r}_{i_2}-{\bf r}_{1})
{\mathcal V}({\bf r}_{1}-{\bf r}_{0})
 \otimes 
 {\mathcal V}^*({\bf r}_{i_{j}}-{\bf r}_{i_{j+1}}) \dots 
{\mathcal V}^*({\bf r}_{i_{n-1}}-{\bf r}_{i_{n}}) 
 {\mathcal V}^*({\bf r}_{i_{n}}-{\bf r}_{0}) \, .
\label{eq:uward1}
\end{eqnarray}

\end{widetext}

As an example, Fig.~\ref{fig:cross-a-b}(b) displays the $U$-diagram obtained from the $\Sigma$-diagram shown  in Fig.~\ref{fig:cross-a-b}(a) by cutting at the impurity $2$. If we try to apply the recipe to a self-energy diagram which is not two-point irreducible we obtain a reducible intensity diagram. For instance, cutting the diagram of Fig.~\ref{fig:cross-c-d}(a) at the impurity 3 (inside the red rectangle) leads to the intensity diagram of Fig.~\ref{fig:cross-c-d}(b), which is not irreducible in the sense defined in Appendix \ref{apprid}. It is not difficult to see that this example can be generalized as follows: 
any two-point reducible $\Sigma$-diagram generates a reducible intensity diagram, and vice versa, any reducible intensity diagram is generated by a $\Sigma$-diagram reducible diagram. Therefore, all irreducible $U$-diagrams (and only them) are generated by two-point irreducible $\Sigma$-diagrams. 

\subsection{Proof of the Ward identity}
\label{appWI}

The above-described construction of the  irreducible component of the intensity propagator ensures the fulfillment of the Ward identity presented in Eq.~(\ref{eq:ward0}). In order to prove such a relationship, we transform \eqref{eq:usumward} to momentum space and take ${\bf q}=0$ for each term $k$ in the sum, obtaining

\begin{widetext}

\begin{eqnarray}
\tilde{U}_{{\bf i},j}^{\sigma_1\sigma_2,\sigma_3\sigma_4}(\varepsilon,\omega,0) & = & n_\mathrm{i}^l \
\left[G^{(+)}\left(\varepsilon_1\right)\right]^{j-1}
\left[G^{(-)}\left(\varepsilon_2\right)\right]^{n-j}
\int{\rm d}{\bf r}_1\dots{\rm d}{\bf r}_l 
\left[
{\mathcal V}({\bf r}_0-{\bf r}_{i_{j-1}}) 
\dots {\mathcal V}({\bf r}_{i_2}-{\bf r}_{1})
{\mathcal V}({\bf r}_{1}-{\bf r}_{0}) 
\right]^{\sigma_1\sigma_3} 
\nonumber\\
& & \left[
 {\mathcal V}^*({\bf r}_{0}-{\bf r}_{i_{j+1}}) \dots 
{\mathcal V}^*({\bf r}_{i_{n-1}}-{\bf r}_{i_{n}}) 
 {\mathcal V}^*({\bf r}_{i_{n}}-{\bf r}_{0})
\right]^{\sigma_2\sigma_4} \, .
\label{eq:uward3}
\end{eqnarray}
The above expression holds for both of the cases ($i_j = 0$ and $i_j\neq 0$). Taking $\sigma_2=\sigma_1$ and summing over the repeated spin index we have

\begin{eqnarray}
\sum_{\sigma_1}
\tilde{U}_{{\bf i},j}^{\sigma_1\sigma_1,\sigma_3\sigma_4}(\epsilon,\omega,0) & = & 
n_\mathrm{i}^l \
\left[G^{(+)}\left(\varepsilon_1\right)\right]^{j-1}
\left[G^{(-)}\left(\varepsilon_2\right)\right]^{n-j}
\int{\rm d}{\bf r}_1\dots{\rm d}{\bf r}_l 
\left[
{\mathcal V}({\bf r}_0-{\bf r}_{i_{n}}) 
\dots
{\mathcal V}({\bf r}_{i_2}-{\bf r}_{1}) {\mathcal V}({\bf r}_{1}-{\bf r}_0) 
\right]^{\sigma_4\sigma_3} \, ,
\label{eq:uward4}
\end{eqnarray}
where we have used the property \eqref{eq:vomrka}.
Notice that the integrals appearing in Eq.~(\ref{eq:uward4}) are exactly those of Eq.~(\ref{eq:sigmaward}). 

We now consider the sequence $\bf{\tilde{i}}$ where the impurities are visited in the reversed order, i.e. $\tilde{i}_j=i_{n+1-j}$. The corresponding self-energy $\Sigma^{(+)}_{{\bf \tilde{i}}}(\varepsilon)$ has the same structure of \eqref{eq:sigmaward}, up to the reverse of the impurity sequence. Therefore,

\begin{equation}
\Sigma^{(-)}_{{\bf \tilde{i}}}(\varepsilon) = 
\left[\Sigma^{(+)}_{{\bf \tilde{i}}}(\varepsilon)\right]^*
  =  n_\mathrm{i}^l \
 \left[G^{(-)}(\varepsilon)\right]^n 
 \int{\rm d}{\bf r}_1\dots{\rm d}{\bf r}_l 
{\mathcal V}^*({\bf r}_0-{\bf r}_{1}) 
{\mathcal V}^*({\bf r}_{1}-{\bf r}_{i_{2}})\dots
{\mathcal V}^*({\bf r}_{i_{n-1}}-{\bf r}_{i_n}) {\mathcal V}^*({\bf r}_{i_n}-{\bf r}_0) \, .
\label{eq:sigmacward}
\end{equation}
The diagonal character of the self-energy ensures that 
$\Sigma_{\bf \tilde{i}}^T=\Sigma_{\bf \tilde{i}}$. Using again Eq.~\eqref{eq:vomrka} we have
\begin{equation}
\Sigma^{(-)}_{{\bf \tilde{i}}}(\varepsilon) = n_\mathrm{i}^l \
 \left[G^{(-)}(\varepsilon)\right]^n  
\int{\rm d}{\bf r}_1\dots{\rm d}{\bf r}_l 
{\mathcal V}({\bf r}_0-{\bf r}_{i_{n}}) 
{\mathcal V}({\bf r}_{i_{n}}-{\bf r}_{i_{n-1}})\dots
{\mathcal V}({\bf r}_{i_2}-{\bf r}_1) {\mathcal V}({\bf r}_{1}-{\bf r}_0) \, .
\label{eq:sigmacward2}
\end{equation}

 The identity
\begin{equation}
\left[G^{(+)}(\varepsilon_1)\right]^n-\left[G^{(-)}(\varepsilon_2)\right]^n  = \bigl[G^{(+)}(\varepsilon_1)-G^{(-)}(\varepsilon_2)\bigr] \sum_{k=1}^n \left[G^{(+)}(\varepsilon_1)\right]^{k-1}\left[G^{(-)}(\varepsilon_2)\right]^{n-k} \, ,
\end{equation}
together with Eqs.~(\ref{eq:sigmaward},\ref{eq:uward4},\ref{eq:sigmacward}), yield:
\begin{equation}
\Sigma_{\bf i}^{\sigma_4\sigma_3(+)}(\varepsilon_1) -\Sigma_{\bf\tilde{i}}^{\sigma_4\sigma_3(-)}(\varepsilon_2) = \bigl[G^{(+)}(\varepsilon_1)-G^{(-)}(\varepsilon_2)\bigr]\sum_{j=1}^{n} \sum_{\sigma_1} \tilde{U}_{{\bf i},j}^{\sigma_1\sigma_2,\sigma_3\sigma_4}(\epsilon,\omega,0) \, .
\label{eq:wardapp}
\end{equation}

\end{widetext}

Thus, Eq.~(\ref{eq:ward0}) holds for each contribution to the self-energy described by a reciprocity-invariant sum of diagrams. The latter are defined by the condition that for each individual contributing sequence diagram ${\bf i}$, its reversed counterpart ${\bf \tilde{i}}$ is also included. The condition ${\bf \tilde{i}}={\bf i}$ (like for instance in the case of self-energy in Fig.~\ref{fig:fig1_GF}(b) constitutes a special case of a reciprocity-invariant diagram. All the three approximations considered in this work are based on reciprocity-invariant diagrams. The equivalence between (\ref{eq:ward0}) and \eqref{eq:wardapp} is ensured by the fact that $G$ and $\Sigma$ are scalar and the sum over $\sigma_2$ carried over in the former equation simply sets $\sigma_2=\sigma_1$.

\begin{widetext}

\section{Explicit expressions of ${\mathcal F}_0^{(\pm)}(\varepsilon,{\bf r})$ and ${\mathcal F}_D^{(\pm)}(\varepsilon,{\bf r})$}
\label{sec:AppendixC}

Identifying the poles of $\tilde{\mathcal F}_0^{(\pm)}(\varepsilon,{\bf k})$ in Eq.~(\ref{eq:F0}), we write
\begin{equation}
\tilde{\mathcal F}_0^{(\pm)}(\varepsilon,{\bf k}) = - \ \frac{32\pi V_0}{a^3
\left(k^2-\left[k_1^{(\pm)}(\varepsilon)\right]^2\right)
\left(k^2-\left[k_2^{(\pm)}(\varepsilon)\right]^2\right)
\left(k^2-\left[k_3^{(\pm)}(\varepsilon)\right]^2\right)}
\, ,
\end{equation}
where $k=|{\bf k}|$ and
\begin{equation}
k_j^{(\pm)}(\varepsilon)=\frac{\sqrt{-1+e^{\pm 2(j-1)i\pi/3}\left(-32\pi a^3 n_\mathrm{i} \ V_0 \ G^{(\pm)}(\varepsilon) \right)^{1/3}}}{a} \, , \ j=1,2,3 \, .
\end{equation}
We note that ${\rm Im}\left[k_{1,2}^{(+)}(\varepsilon)\right]>0$ and ${\rm Im}\left[k_{3}^{(+)}(\varepsilon)\right]<0$, whereas the opposite signs hold for $k_j^{(-)}(\varepsilon)$.
The inverse Fourier transform can be written as:
\begin{equation}
{\mathcal F}_0^{(\pm)}(\varepsilon,{\bf r}) = \int
\frac{{\rm d}{\bf k}} {(2\pi)^3} e^{-i{\bf k}\cdot {\bf r}} \ \tilde{\mathcal F}_0^{(\pm)}(\varepsilon,{\bf k}) =
\frac{1}{4\pi^2 r}\int_{-\infty}^\infty {\rm d}k~k\sin(k r) \ \tilde{\mathcal F}_0^{(\pm)}(\varepsilon,k)
\end{equation}
The integral over $k$ can now be performed using residual calculus. The result simplifies if it is multiplied by
\begin{equation}
G^{(\pm)}(\varepsilon)=\pm \frac{
i a^3 
\left(
\left[k_1^{(\pm)}(\varepsilon)\right]^2-
\left[k_2^{(\pm)}(\varepsilon)\right]^2
\right)
\left(
\left[k_1^{(\pm)}(\varepsilon)\right]^2-
\left[k_3^{(\pm)}(\varepsilon)\right]^2
\right)
\left(
\left[k_2^{(\pm)}(\varepsilon)\right]^2-
\left[k_3^{(\pm)}(\varepsilon)\right]^2
\right)
}{
96\sqrt{3} n_\mathrm{i} \pi V_0
} \, ,
\end{equation}
and it then reads:
\begin{equation}
{\mathcal F}_0^{(\pm)}(\varepsilon,{\bf r}) G^{(\pm)}(\epsilon) = \pm i \ \frac{\left(e^{\pm ik_1^{(\pm)}(\varepsilon) r}-e^{\pm ik_2^{(\pm)}(\varepsilon) r}\right)\left[k_3^{(\pm)}(\varepsilon)\right]^2+\left(e^{\pm ik_2^{(\pm)}(\varepsilon) r}-e^{\mp ik_3^{(\pm)}(\varepsilon) r}\right)\left[k_1^{(\pm)}(\varepsilon)\right]^2+\left(e^{\mp ik_3^{(\pm)}(\varepsilon) r}-e^{\pm ik_1^{(\pm)}(\varepsilon) r}\right)\left[k_2^{(\pm)}(\varepsilon)\right]^2}{12 \sqrt{3} \pi  n_\mathrm{i} r} \, .
\label{eq:f0r}
\end{equation}

From the definition \eqref{eq:FD} we must calculate ${\mathcal F}_D^{(\pm)}(\varepsilon,{\bf r})$ as an inverse Fourier transform. Performing the angular part of the integration over ${\bf k}$ we obtain
\begin{equation}
{\mathcal F}_D^{(\pm)}(\varepsilon,{\bf r})=\frac{16 \gamma \left[(x\pm i y) x y-(y\pm i x)z^2\right]}{\pi a^9 r^7}
\int_{-\infty}^\infty {\rm d}k  \frac{k(1+a^2 k^2)^2 \left(3 (5-2 k^2 r^2)\sin(k r)-k r (15-k^2 r ^2)\cos(k r)\right)
}
{
\left(k^2-\left[k_1^{(\pm)}(\varepsilon)\right]^2\right)^2
\left(k^2-\left[k_2^{(\pm)}(\varepsilon)\right]^2\right)^2
\left(k^2-\left[k_3^{(\pm)}(\varepsilon)\right]^2\right)^2} \, .
\end{equation}
\end{widetext}
This integral can again be performed using residual calculus. However, since the denominator now contains second-order poles, the result is considerably more complicated than Eq.~(\ref{eq:f0r}), and we refrain from displaying it here.


\begin{thebibliography}{99}

\bibitem{cha} 
J.-N.\ Chazalviel, 
Phys.\ Rev.\ B \textbf{11}, 1555 (1975).

\bibitem{zar-cas}
V.\ Zarifis and T.\ G.\ Castner, 
Phys.\ Rev.\ B \textbf{36}, 6198 (1987).

\bibitem{kik-aws}
J.\ M.\ Kikkawa and D.\ D.\ Awschalom,
Phys.\ Rev.\ Lett.\ \textbf{80}, 4313 (1998).

\bibitem{dzh}
R.\ I.\ Dzhioev {\it et al.}, 
Phys.\ Rev.\ B \textbf{66}, 245204 (2002).

\bibitem{oes-roe-hau-hae}
M.\ Oestreich,  M. R\"omer, R.J. Haug, and D. H\"agele, 
Phys.\ Rev.\ Lett.\ \textbf{95}, 216603 (2005).

\bibitem{sch-hei-roh}
L.\ Schreiber, M. Heidkamp, T. Rohleder, B. Beschoten, and G. G\"untherodt,  arXiv:0706.1884v1 (2007).

\bibitem{roe-ber-mue} M.\ R\"omer, H. Bernien, G. M\"uller, D. Schuh, J. H\"ubner, and M. Oestreich,
Phys.\ Rev.\ B \textbf{81}, 075216 (2010).

\bibitem{spr-etal}
D.\ Sprinzl {\it et al.}, 
Phys.\ Phys.\ B \textbf{82}, 153201 (2010).

\bibitem{fab-etal}
J.\ Fabian, A. Matos-Abiaguea, C. Ertlera, P. Stano, and I. \u{Z}utic,
Acta Physica Slovaca \textbf{57}, 565 (2007).

\bibitem{wu-jia-wen}
M.W.\ Wu, J.H.\ Jiang, and M.Q.\ Weng,
Physics Reports {\bf 493}, 61 (2010). 

\bibitem{vLoe}
H. v. L\"ohneysen, 
Current Opinion in Solid State \& Material Science {\bf 3}, 5 (1988).

\bibitem{gehard}
F. Gebhard, 
{\it The Mott Metal Insulator Transition: Models and Methods}, 
Springer (Berlin, 2010).

\bibitem{slevin14}
Y. Harashima and K. Slevin,
Phys.\ Rev.\ B {\bf 89}, 205108 (2014).

\bibitem{lau01}
W.\ H.\ Lau, J.\ T.\ Olesberg, and M.\ E.\ Flatt\'{e},
Phys.\ Rev.\ B \textbf{64}, 161301(R) (2001).

\bibitem{jia-wu}
J.\ H.\ Jiang and M.\ W.\ Wu, 
Phys.\ Rev.\ B {\bf 79}, 125206 (2009).

\bibitem{shk} 
B.\ I.\ Shklovskii, 
Phys.\ Rev.\ B \textbf{73}, 193201 (2006).

\bibitem{kav} 
K.\ V.\ Kavokin, 
Phys.\ Rev.\ B \textbf{64}, 075305 (2001).

\bibitem{put-joy}
W.\ O.\ Putikka and R.\ Joynt, 
Phys.\ Rev.\ B \textbf{70}, 113201 (2004).

\bibitem{tam-wei-jal} 
P.\ I.\ Tamborenea, D.\ Weinmann, and R.\ A.\ Jalabert, 
Phys.\ Rev.\ B \textbf{76}, 085209 (2007).

\bibitem{prl2012}
G.\ A. Intronati, P.\ I.\ Tamborenea, D.\ Weinmann, and R.\ A.\ Jalabert, 
Phys.\ Rev.\ Lett.\ \textbf{108}, 016601 (2012).

\bibitem{twj2016}
G.\ A. Intronati, P.\ I.\ Tamborenea, D.\ Weinmann, and R.\ A.\ Jalabert, 
unpublished (2016).

\bibitem{car-chr-fas} 
M.\ Cardona, N.\ E.\ Christensen, and G.\ Fasol, 
Phys.\ Rev.\ B \textbf{38}, 1806 (1988).

\bibitem{eng-ras-hal}
H-A.\ Engel, E.\ I.\ Rashba, and B.\ I.\ Halperin, in
{\it Handbook of Magnetism and Advanced Magnetic Materials},
Vol.\ 5, H.\ Kronm\"uller and S.\ Parkin (eds.)
(John Wiley \& Sons Ltd, Chichester, 2007).

\bibitem{win} 
{\it Spin-orbit coupling effects in two-dimensional electron and hole systems}, 
R.\ Winkler (Springer-Verlag, Berlin, 2003). 

\bibitem{mat-toy}
T.\ Matsubara and Y.\ Toyozawa, 
Prog.\ Theoret.\ Phys.\ \textbf{26}, 739 (1961).

\bibitem{proceedings}
G.\ A. Intronati, P.\ I.\ Tamborenea, D.\ Weinmann, and R.\ A.\ Jalabert, 
Physica B \ \textbf{407}, 3252 (2012).

\bibitem{Elyutin}
P.\ V. Elyutin,
J. Phys. C: Solid State Phys. \textbf{14}, 1435 (1981).

\bibitem{Shklovskii}
B.I. Shklovskii, A.L. Efros, {\it Electronic Properties of Doped Semiconductors}, Springer Verlag (New York, Tokyo, 1984).

\bibitem{Economou}
E.N. Economou, A.C. Fertis, in {\it Localization and Metal-Insulator Transitions}, edited by H. Fritzsche and D. Adler, Plenum Press (New York, 1985).

\bibitem{Romero}
D. Romero, S. Liu, H.D. Drew, and K. Ploog
Phys.\ Rev.\ B \textbf{42}, 3179 (1990).

\bibitem{yonezawa64}
F.\ Yonezawa, 
Prog.\ Theoret.\ Phys.\ \textbf{31}, 357 (1964).

\bibitem{mat-kan}
T.\ Matsubara and T.\ Kaneyoshi, 
Prog.\ Theoret.\ Phys.\ \textbf{36}, 695 (1966).

\bibitem{gas-cyr}
F. Cyrot-Lackmann and J.P. Gaspard , 
J. Phys. C: Solid State Phys. \textbf{7}, 1829 (1974).

\bibitem{chao-oli-majlis}
K.A. Chao, F.A. Oliveira, and N. Majlis,
Solid State Communications \textbf{21}, 845 (1977). 

\bibitem{pur-oda_81} 
A.\ Puri and T.\ Odagaki, 
Phys.\ Rev. B {\bf 24},5541 (1981).

\bibitem{gibbons_81} 
M.K.\ Gibbons, D.E.\ Logan, and P.A.\ Madden, 
Phys.\ Rev. B {\bf 38}, 7292 (1988).

\bibitem{chi-hub} 
W.Y.\ Ching and D.L.\ Huber, 
Phys.\ Rev.\ B {\bf 26}, 5596 (1982).

\bibitem{Rydberg_ex}
G. G\"unter {\it et al.}, 
Science \textbf{342}, 954 (2013). 

\bibitem{Rydberg_th}
T. Scholak, T. Wellens, and A. Buchleitner,
Phys.\ Rev.\ A \textbf{90}, 063415 (2014).

\bibitem{amerongen00}
H. van Amerongen, L. Valkunas, and R. van Grondelle, 
{\em Photosynthetic Excitons}, World Scientific (Singapore, 2000).

\bibitem{scholak11}
T. Scholak, T. Wellens, and A. Buchleitner,
Europhys. Lett. \textbf{96}, 10001 (2011). 

\bibitem{engel07}
G. Engel {\it et. al.}, 
Nature {\textbf 446}, 782 (2007).

\bibitem{noz-lew} 
P.\ Nozi\`eres and C.\ Lewiner,
J.\ Phys.\ (Paris) \textbf{34}, 901 (1973).

\bibitem{cha-van-kot} 
A.N.\ Chantis, M.\ van Schilfgaarde, and T.\ Kotani, 
Phys.\ Rev.\ Lett.\ \textbf{96}, 086405 (2006).

\bibitem{kri-hal}                                          
J.J.\ Krich and B.I.\ Halperin, 
Phys.\ Rev.\ Lett.\ \textbf{98}, 226802 (2007).

\bibitem{mar-ste-tit} 
V.\ A.\ Maruschek, N.\ M.\ Stepanova, and A.\ N.\ Titkov, 
Fiz.\ Tervd.\ Tela (Donelsk) \textbf{215}, 3537 (1983).

\bibitem{jusserand}
B.\ Jusserand, D. Richards, G. Allan, C. Priester, and B. Etienne,
Phys.\ Rev.\ B \textbf{51}, 4707 (1995).

\bibitem{knap-etal} 
W.\ Knap {\it et al.},
Phys.\ Rev.\ B \textbf{53}, 3912 (1996).

\bibitem{mei-etal}
L.\ Meier {\it et al.},
Nature Phys.\ \textbf{3}, 650 (2007).

\bibitem{Serre}
J. Serre and A. Ghazali,
Phys. Rev. B \textbf{28}, 4704 (1983).

\bibitem{hal-lax}                                          
B.I.\ Halperin and M. Lax 
Phys.\ Rev.\ \textbf{148}, 772 (1966).

\bibitem{vol-woe}
D. Vollhardt and P. W\"olfle,
Phys.\ Rev.\ B \textbf{22}, 4666 (1980).

\bibitem{rev-vol-woe}
D. Vollhardt and P. W\"olfle,
in {\it Electronic phase transitions}, edited by W. Hanke and Yu. V. Kopaev (North-Holland, Amsterdam 1992), p. 1.

\bibitem{akk-mont}
E. Akkermans and G. Montambaux
{\it Mesoscopic Physics of Electrons and Photons},
Cambridge University Press (New York, 2007).

\bibitem{Radjenovic}
B. Radjenovic and D. Tjapkin,
Phys. Stat. Sol. (b) \textbf{166}, 487 (1989).

\bibitem{kkw90}
J. Kroha, T. Kopp, and P. W\"olfle, 
Phys.\ Rev.\ B \textbf{41}, 888 (1990).

\bibitem{schneider10}
M. Krau\ss, H. C. Schneider, R. Bratschitsch, Z. Chen, and S. T. Cundiff,
Phys.\ Rev.\ B \textbf{81}, 035213 (2010).

\bibitem{beschoten01}
B. Beschoten {\it et al.},
Phys.\ Rev.\ B \textbf{63}, 121202(R) (2001).

\bibitem{edwards} 
S.F. Edwards, Philos. Mag., \ \textbf{3}, 1020 (1958).

\bibitem{Leath}
P.L. Leath, 
Phys. Rev. B \textbf{2}, 3078 (1970).

\bibitem{altshuler82}
B.L.Altshuler, A.G.Aronov, and A.Yu. Zuzin,
Solid State Comm., \textbf{44},137 (1982).

\bibitem{sachdev87}
S. Sachdev,
Phys. Rev. B \textbf{35}, 7558 (1987).

\bibitem{raimondi90}
R. Raimondi, C. Castellani, and C. Di Castro,
Phys. Rev. B \textbf{42}, 4724 (1990).

\bibitem{geiger12}
T. Geiger, T. Wellens, and A. Buchleitner, 
Phys. Rev. Lett. {\bf 109}, 030601 (2012).

\bibitem{geiger13}
T. Geiger, A. Buchleitner, and T. Wellens, 
New J. Phys. {\bf 15}, 115015 (2013). 

\end{thebibliography}
\end{document}